\documentclass[a4paper,12pt]{article}
\usepackage{amsfonts,amsthm,amsmath,amssymb,graphicx,hyperref,color}

\title{Diagonal multi-matrix correlators and BPS operators in $\cN=4$ SYM}

\def\bra#1{\left\langle #1\right|}
\def\ket#1{\left| #1\right\rangle}
\newcommand{\braket}[2]{\langle #1 | #2 \rangle}

\def\corr#1{\left\langle #1 \right\rangle}
\newcommand{\tr}{\operatorname{tr}}
\newcommand{\Str}{\operatorname{Str}}

\newcommand{\Sym}{\operatorname{Sym}}

\def\id{\textrm{id}}

\def\Dim{\textrm{Dim}}
\def\End{\textrm{End}}

\def\Sym{\textrm{Sym}}

\def\a{\alpha}
\def\b{\beta}

\def\g{\gamma}
\def\G{\Gamma}

\def\m{\mu}

\def\r{\rho}

\def\s{\sigma}

\def\l{\lambda}
\def\L{\Lambda}

\def\D{\Delta}

\def\tA{ {\tilde A} }

\newcommand{\cN}{\mathcal N}
\newcommand{\cO}{\mathcal O}

\newcommand{\cZ}{\mathcal Z}

\newcommand{\be}{\begin{equation}}
\newcommand{\bea}{\begin{eqnarray}}

\newcommand{\ee}{\end{equation}}
\newcommand{\eea}{\end{eqnarray}}

\newcommand{\nn}{\nonumber}

\begin{document}

\rightline{QMUL-PH-07-23}

\vspace{2truecm}

\vspace{15pt}

%%%%%%%%%%%%%%%%%

\centerline{\LARGE \bf Diagonal multi-matrix correlators } 
\centerline{ \LARGE \bf  and BPS operators in $\cN=4$ SYM } \vspace{1truecm}
\thispagestyle{empty} \centerline{
    {\large \bf T.W. Brown}\footnote{E-mail address:
                                  {\tt t.w.brown@qmul.ac.uk}},
    {\large \bf P.J. Heslop}\footnote{E-mail address:
                                  {\tt p.j.heslop@qmul.ac.uk}}
    {\bf and}
    {\large \bf S. Ramgoolam}\footnote{E-mail address:
                                  {\tt s.ramgoolam@qmul.ac.uk}}
                                                       }

\vspace{.4cm}
\centerline{{\it  Centre for Research in String Theory, Department of Physics}}
\centerline{{ \it Queen Mary, University of London}}
 \centerline{{\it Mile End Road, London E1 4NS, UK}}

\vspace{1.5truecm}

%%%%%%%%%%%%%%%%%
\thispagestyle{empty}

\centerline{\bf ABSTRACT}

\vspace{.5truecm}

\noindent We present a complete basis of  multi-trace multi-matrix operators
that has a diagonal two point function for the free matrix field theory at
finite $N$.  This generalises to multiple matrices the single matrix
diagonalisation by  Schur polynomials.  Crucially, it involves
intertwining the gauge group $U(N)$ and the global symmetry group
$U(M)$ with Clebsch-Gordan coefficients of symmetric groups $S_n$.
When applied to $\cN = 4$ super Yang-Mills we consider the $U(3)$
subgroup of the full symmetry group.  The diagonalisation allows the
description of a dual basis to multi-traces, which permits the
characterisation of the metric on operators transforming in short
representations at weak coupling. This gives a framework
for the comparison of quarter and  eighth-BPS giant
gravitons of $ AdS_5 \times S^5 $ spacetime to gauge invariant 
operators of the dual $\cN = 4$ SYM.

\vspace{.5cm}

\newpage

\tableofcontents

\setcounter{footnote}{0}

%---------------------------------------------------------------------------

\section{Introduction}

$\cN=4$ super-Yang Mills with $U(N)$ gauge group has six $ N \times N
$ hermitian scalar matrices which can be organised into three complex
scalar matrices. At zero Yang Mills coupling, the gauge invariant
holomorphic functions of the matrices give rise to $1/8$ BPS operators
and by the operator-state correspondence, $1/8$ BPS states. These
states have a metric determined by the two-point function. The space
of holomorphic operators can be organised into representations of a
$U(3)$ subgroup of the $SO(6)$ $R$-symmetry. When we consider the
analogous problem with a single complex matrix, we have $U(1)$
symmetry, and the operators are half-BPS. The diagonalisation of the
two-point function is accomplished by a basis of Schur polynomials
\cite{cjr}.  Thanks to the AdS/CFT duality \cite{malda}, this allows a
map from gauge theory states to giant gravitons \cite{mst,giantgravitons}
and LLM \cite{Lin:2004nb} geometries in AdS-spacetime. In this paper
we will solve the analogous diagonalisation problem for the  $1/4$ and 
$1/8$-BPS operators. The $1/4$-BPS case involves a $U(2)$ subgroup of 
the $U(3)$. 

We will consider, in general, 
 a $U(N)$ gauge theory with global $U(M)$ symmetry.
We have complex scalar fields which are valued in the adjoint
representation of the gauge group $U(N)$. The fields $(X_{a})^i_j$
(where $a$ is a fundamental $U(M)$ index and $i$ and $j$ are 
$U(N)$ indices) have a free field correlator with index structure
\begin{equation}\label{basiccorrelatorUN}
  \corr{ (X_{a})^i_j (X_{b}^\dagger)^k_l} = \delta_{ab}\, \delta^i_l \delta^k_j
\end{equation}
We have ignored the  trivial spacetime dependence $ |x_1 - x_2|^{-2} $ 
of the correlator, which is determined by conformal symmetry.
While our primary interest
is in the four-dimensional gauge theory, our main results have to do
with the colour and flavour structure of the operators. Hence they are
equally applicable to the reduced matrix models in one or zero
dimension as long as the correlator (\ref{basiccorrelatorUN}) is
valid.

We can build gauge-invariant polynomials in these fields by taking
traces over the $U(N)$ indices of polynomials in the $X_{a}$.  For
example if $M=2$ we have fields $X$ and $Y$ and we can build operators
such as $\tr(XXY)$ and $\tr(XX)\tr(Y)$.  In this paper, we provide a
complete linearly independent basis  that spans 
these multi-trace multi-field operators. 
It has the correct counting of multi-trace multi-field
operators given by P\'olya theory.  Furthermore it diagonalises the
metric at zero coupling for finite $N$.

\subsection{Summary of main results}

We give here a summary of our key results.  We find that a complete
diagonal basis of multi-trace multi-matrix operators is given by
operators of the form \eqref{simplifiedop}:
\begin{equation} 
   \cO^{ \L \mu , R }_{ \b, \tau }= \frac{1}{n!}\sum_{\a} B_{j \b} \; S^{\tau  ,}{}^{  \L }_{  j
  }\;{}^{R}_{p }\;{}^{R}_{q}\;\; D_{pq}^R(\a) \tr(\a\;
  {\bf X}^{\mu} )
\nn
\end{equation}
The $D^R_{pq} $ in this equation are matrix elements of the symmetric
group on $n$ letters, in the irreducible representation (irrep.) $R$.
$R$ also labels an irrep. of $U(N)$; $\L$ labels an irrep. of $U(M)$.
${\bf X^{\mu}}$ is an abbreviated notation for a tensor product of $n$
matrix fields and the trace is being taken in $V^{\otimes n }$, where
$V$ is an $N$-dimensional vector space, the fundamental representation
of $U(N)$.  The $S$ factor is a Clebsch-Gordan coefficient coupling a
tensor product of irreducible representations $ R \otimes R $ of the
symmetric group to the irrep.  $ \Lambda $. The $B$-factor is a
branching coefficient for the restriction of the representation $
\Lambda $ of the symmetric group $S_n$ to the trivial one-dimensional
irrep. of its subgroup $S_{\mu_1 } \times S_{\mu_2 } \cdots \times
S_{\mu_M } $. In the case at hand we are considering three complex
matrices so $M=3$. $\mu_1 , \mu_2 , \mu_3 $ give the numbers of the 3
complex matrices in the operator.  The branching coefficients are
explained in more detail in Section \ref{sec:opcor}. Useful formulae
on relations between matrix elements and Clebsch-Gordan coefficients
are collected in Appendix \ref{sec:formulae}.  The proof that the
operators (\ref{simplifiedop}) provide a complete set of gauge
invariant multi-matrix operators is given at the end of Section
\ref{sec:invrecovertrce}. The diagonality of the two-point function
\begin{align} 
 \corr{ \cO^{\L_1 \mu^{(1)} ,R_1}_{ \beta_1,\tau_1 } \cO^\dagger{}^{\L_2 \mu^{(2)}  ,R_2}_{ \beta_2 ,
  \tau_2 } } = 
\delta^{ \m^{(1)} \m^{(2)} }
\delta^{{\L}_1 {\L}_2 }  \delta_{ \beta_1 \beta_2 }\delta^{R_1 R_2 } \delta_{ \tau_1 \tau_2 }
  \frac{ |H_{\m^{(1)}}|  \Dim R_1}{d_{R_1}^2}\label{orthoresult} 
\nn 
\end{align} 
is derived as \eqref{orthoresult}  in section \ref{sec:ginvdiagbas}.  
  $ |H_\m| $ is the dimension $ \mu_1! \mu_2! \cdots \mu_M! $ 
of the group $H_\m \equiv S_{\mu_1 } \times S_{\mu_2 } \cdots \times
S_{\mu_M } $.

As a prelude to the discussion of gauge invariant operators in Section
\ref{sec:invariant}, we consider in Section \ref{sec:covariant} the
counting and two-point functions of untraced covariant operators.
The mathematics of Schur-Weyl duality which relates representations 
of symmetric groups to those of unitary groups appears repeatedly 
in this paper. Section \ref{sec:covariant}  uses the relation between 
 irreps $\L $ of $S_n$, associated to a  Young diagram $\L$ with 
$n$ boxes,  and the corresponding irrep of $U(M)$. Section 
\ref{sec:invariant}  uses the relation between irreps $R $
of $S_n$ and the corresponding irrep of $U(N)$.
Section \ref{sec:fermions} describes the extension to fermions, 
where the $U(M)$ global symmetry is replaced by $U(M_1|M_2)$. 
Section \ref{extcorr} describes the generalisation from two-point 
to multi-point correlators. We focus on extremal correlators which 
have non-renormalisation properties due to supersymmetry. 
  Section \ref{sec:finiteNproj} gives a general
description of the finite $N$ projector on multi-matrix
multi-traces. While the diagonality \eqref{orthoresult}
is derived at zero-coupling, thanks to supersymmetry, it has 
implications for weak and strong coupling physics. 
 Section \ref{sec:physics} describes 
 the disentangling of operators which are in long
representations at weak coupling from those that are genuine
 non-renormalised 1/8-BPS operators, and a characterisation of the
metric on these operators.    This allows a discussion of
relations to the physics of giant gravitons at strong coupling and the
related harmonic oscillator system.  Since our key results depend only
on the basic formula \eqref{basiccorrelatorUN}, they also apply to
reduced matrix theories of multiple matrix fields in lower dimensions,
such as $0$ or $1$.  In the case of the reduced quantum mechanics
considered  in Section \ref{sec:redmat}, i.e $1$-dimensional reduction, 
we discuss  multiple commuting Hamiltonians related to the 
diagonal basis.

\section{Gauge-covariant operators}\label{sec:covariant}

Consider tensor products of operators of the form 
\begin{equation}\label{endtensn}  
 \hat \cO  (  \vec a ) \equiv
  X_{a_1}  \otimes X_{a_2} \otimes \cdots \otimes X_{a_n } 
\end{equation}
forming words of length $n$.  The $a_i$ can take values from $1$ to
$M$.  The operators correspond to states in $V_M^{\otimes n}$ where
$V_M$ is the fundamental representation of $U(M)$; this dictates the
action of $U(M)$ on this operator. Their number is $M^n$.

There is also an action of the permutation group $\s \in S_n$ on these
operators given by re-ordering
\begin{equation}\label{eq:snactiononstate}
  X_{a_{\s^{-1}(1)}}  \otimes X_{a_{\s^{-1}(2)}} \otimes \cdots \otimes
  X_{a_{\s^{-1}(n)} } 
\end{equation}
If we recall that $X_{a} $ is an $N \times N $ matrix, i.e a linear
transformation of an N-dimensional space $V$, then the word
(\ref{endtensn}) is a linear transformation of $V^{\otimes n }$.  It
follows that the permutation of operators can be achieved by a
conjugation
\begin{equation}\label{eq:snactiononstateI}
  X_{a_{\s^{-1}(1)}}  \otimes X_{a_{\s^{-1}(2)}} \otimes \cdots \otimes
  X_{a_{\s^{-1}(n)} } =  \s \cdot X_{a_1}  \otimes X_{a_2} \otimes
  \cdots \otimes X_{a_n } \cdot \s^{-1}
\end{equation}
where the permutations are now acting on the $n$ factors in $V^{\otimes n } $. 
In terms of indices, permuting the operators is equivalent to 
a simultaneous permutation of  upper and lower indices. In the remainder 
of this section, we will be primarily interested in the $U(M)\times S_n$ 
action on the operators.

Since the $U(M)$ and $S_n$ actions commute we have by Schur-Weyl
duality  (see for example \cite{zelobenko,fultonharris})  
\begin{align}\label{schwey} 
 V_M^{\otimes n } = \oplus_{\Lambda} V_{\Lambda}^{U(M)}  \otimes
 V_{\Lambda} ^{S_n }   
\end{align}
The sum here is over Young diagrams $\L$ with $n$ boxes and at most
$M$ rows that correspond to representations both of $U(M)$ and $S_n$.

The content of (\ref{schwey}) is that there exists an invertible
transformation $C_{a_1 \cdots a_n }^{{\Lambda} , m , i }$ from the
multi-index tensors to the irreducible representations of $ U(M)
\times S_n $.
\begin{align}\label{mapfancytobasic}
X_{a_1}  \otimes X_{a_2} \otimes \cdots \otimes X_{a_n } = \sum_{ {\Lambda}  \vdash n } \sum_{i=1 }^{d_{\Lambda}  } 
\sum_{m = 1}^{\Dim_M  {\Lambda}  }   
  C_{a_1 \cdots a_n }^{{\Lambda}  ; ~  i , m } ~ \hat{\cO}^\L_{i,m}
\end{align} 
Here $i$ labels the $d_\L$ states of the $S_n$ representation $\L$ and
$m$ labels the $\Dim_M \L$ states of the $U(M)$ representation $\L$.

The inverse is 
\begin{align}\label{mapbasictofancy}
\hat{\cO}^\L_{i,m} = \sum_{ a_1 ... a_n } C_{{\Lambda} ; ~  i , m
}^{a_1, .. , a_n } ~  X_{a_1}  \otimes X_{a_2} \otimes \cdots \otimes X_{a_n }
\end{align}

\subsection{Counting}\label{sec:counting} 

Consider   $ \hat \cO (\vec a  ) $ 
as in   (\ref{endtensn}),   with $a_1, \cdots a_n$ 
chosen such that we have a   fixed field content,
i.e. fixed numbers of $X_1, X_2 \dots X_M$ given by $ \mu_1 , \mu_2
\dots \mu_M$. We will denote this set of integers as $\mu$. 
Such operators can be written as 
\begin{equation}\label{exampleop}
\hat \cO^\m(\sigma)   \equiv    \s 
  \cdot X_{1}^{\otimes \m_1}  \otimes X_{2}^{\otimes \m_2} \otimes
  \cdots \otimes X_{M }^{\otimes \m_M} \cdot \s^{-1}
 = \s \; {\bf X}^{\mu} \; \s^{-1}  
\end{equation}
for some $\s \in S_n$. We have used the abbreviation 
 $ {\bf X}^{\mu} $ for $X_{1}^{\otimes \m_1}  \otimes
 X_{2}^{\otimes \m_2} \otimes
  \cdots \otimes X_{M }^{\otimes \m_M}$.  
Fixing $ \vec a $ to take values from $1 \cdots  M $ determines 
$ \mu $ and $ \sigma $.  
The number of operators with fixed $\mu$ is 
\begin{align}\label{cntfxdcontent}  
\frac{ n!}{ \mu_1 ! \mu_2 ! \cdots \mu_M ! } 
\end{align} 
which is the number of ways of choosing $n$ objects, with $\mu_1 $ of
one kind, $\mu_2 $ of a second kind, and so on up to $ \mu_M $ of the
$M$'th kind.  This is also  the coefficient of $x_1^{\m_1} \cdots
x_M^{\m_M}$ in the polynomial $(x_1 + \cdots + x_M)^n$.  From a group
theory perspective we are counting operators \eqref{exampleop} up to
the symmetry
\begin{equation}
  \s \; {\bf X }^\m \; \s^{-1} \to \s h\; {\bf X }^\m \; h^{-1}\s^{-1}
\end{equation}
where the action of $h \in H_\m = S_{\mu_1 } \times S_{\mu_2} \cdots
S_{\mu_M}$ leaves $ {\bf X }^\m $ unchanged.  Thus we should quotient
on the right by $H_\m$ and count elements in the quotient group $S_n /
H_\m$ which has size
\begin{equation}
  |S_n / H_\m| = \frac{|S_n|}{|H_\m|} = \frac{ n!}{ \mu_1 ! \mu_2 ! \cdots \mu_M ! } 
\end{equation}
The words of fixed content $\mu $ form a vector space of dimension $
\frac{|S_n|}{|H_\m|} $.
Now we want to understand the relation between the choice of $\mu $
and the decomposition (\ref{schwey}) into Young diagrams. 
 What is the dimension of the space of tensors of
fixed ${\Lambda} $ when we restrict the content to be specified by
$\mu $ ? Equivalently this is the number of states  in
 $V_{\Lambda}^{U(M)} \otimes V_{\Lambda} ^{S_n } $ which 
 transform in the irrep. $\mu$ of the $U(1)^M$ subgroup of $U(M)$.

 For
a fixed irrep $V_{\Lambda}^{U(M)} $ this number is
\begin{equation}\label{eq:LRstate} 
 g ( [\mu_1],[\mu_2], \dots [\mu_M]  ; \Lambda )  \equiv g(\m;\L)
\end{equation}
This is  the Littlewood-Richardson
 coefficient for the appearance of
$\L$ in the tensor product of trivial single-row representations of
$U(M)$ $[\mu_1]\otimes \dots \otimes[\mu_M]$.  
This is also known as the
Kostka number and counts the number of $U(M)$ states of $\L$ with
field content $\m$. 
Since the irrep $V_{\Lambda}^{U(M)}$ occurs in tensor space with
multiplicity $ d_{\Lambda } $, we have $ \sum_{\Lambda} g(\m;\Lambda)
 d_{\Lambda }
$ states in $ V_{M}^{\otimes n }$ with fixed content $\mu $. A
combinatoric implication is
\begin{align} 
  { n! \over \mu_1 ! \cdots \mu_M! } 
= \sum_{\Lambda } d_{\Lambda } g( \mu;\Lambda  ) = { |S_n | \over |H_\m| } 
\end{align} 
This is an identity which follows from properties of symmetric group
representations.

\subsection{Operators and correlators}\label{sec:opcor}

Now we are going to implement the map \eqref{mapbasictofancy} that
takes us from $V_M^{\otimes n }$ to a state in $ V_{\Lambda}^{U(M)}
\otimes V_{\Lambda} ^{S_n }$ in order to write down a linearly
independent basis of covariant operators.  To understand where this
basis comes from, we consider the two point function.

To work out the two point function we must reintroduce the $U(N)$
indices.  Each $X_{a_i}$ is a complex matrix in $\End ( V ) $, where
$V$ is an $N$-dimensional vector space, the fundamental of $U(N)$.
Words of length $n$ are elements of $ \End ( V^{\otimes n } )$.  We
write the $U(N)$ indices compactly as
\begin{equation}
  (X_{a_1})^{i_1}_{j_1} \otimes \cdots \otimes (X_{a_n})^{i_n}_{j_n} =
  {\bf X }^{I}_{J}
\end{equation}
Using the $U(N)$ correlator \eqref{basiccorrelatorUN} we get for the
operator ${\bf X}^\m = X_{1}^{\otimes \m_1} \otimes \cdots \otimes
X_{M }^{\otimes \m_M}$
\begin{equation}\label{eq:basicmultifieldcorr}
  {\langle}  ({\bf X }^\mu)^I_J 
({ \bf  X}^\dagger {}^\mu )^{K}_L   {\rangle}= \sum_{ \gamma \in H_{\mu } } \;( \gamma
)^{I}_{L} 
  ( \gamma^{-1} )_{J}^{K} 
\end{equation}
The sum over $\g \in H_{\mu} = S_{\mu_1 } \times \cdots \times
S_{\mu_M }$ is over all the possible ways of contracting the $X_1$'s,
$X_2$'s, etc. of the two operators.

\begin{figure}[t]
\begin{center}
\includegraphics{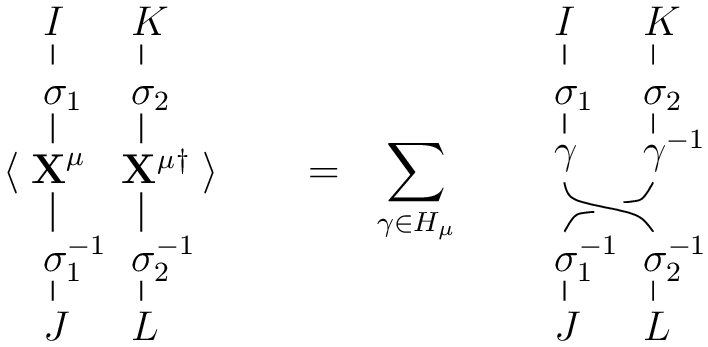}
\caption{two-point function for covariant operator} \label{fig:twopointfunctioncov}
\end{center}
\end{figure}

Next consider the operator \eqref{exampleop}.   
Following the basic correlator \eqref{eq:basicmultifieldcorr} we find
(see Figure \ref{fig:twopointfunctioncov})
\begin{align}
  \corr{(\hat \cO^\m(\sigma_1))^I_J\;\;
  (\hat\cO^\dagger{}^\m(\sigma_2))^K_L} & = \corr{  (\sigma_1 {\bf X }^\mu  \sigma_1^{-1})^I_J 
(\sigma_2 { \bf  X }^\mu  \sigma_2^{-1} )^{K}_L } \nn \\
  & =  \sum_{ \gamma \in H_{\mu } } 
  ( \sigma_1   \gamma \sigma_2^{-1}  )^{I}_{L}
  ( \sigma_2 \gamma^{-1} \sigma_1^{-1} )_{J}^{K} 
\end{align}

It is useful to Fourier expand the group
algebra in terms of matrix elements of irreducible representations,
which are known to give a complete set of functions on the group by
the Peter-Weyl theorem
\begin{align} \label{PW}
\hat \cO^{\Lambda \m}_{ij} = \frac{1}{n!}\sum_{\sigma } D^{ \L }_{ij}
( \sigma ) \hat \cO^\m(\sigma )
\end{align} 
The sum $\sigma$ is over all elements in $S_n$, and we are 
relating operators labelled by $\sigma $ to operators 
labelled by $ ( \L , i ,j ) $.   
Here $D^{ \L }_{ij} ( \sigma )$ is a representing matrix of the $S_n$
representation $\L$. $i$ and $j$ range over the states of this
representation, going from 1 to the dimension $d_\L$.  In particular
we choose  orthogonal representing matrices obeying 
\begin{equation}
  D^{ \L }_{ij} ( \sigma^{-1} ) = D^{ \L }_{ji} ( \sigma )
\end{equation}
such as the ones constructed in Hamermesh~\cite{hamermesh}. 

The two point function of the Fourier-transformed operators ${\hat
\cO}^{\Lambda \m}_{i j}$ is
\begin{align}
  \corr{  ( \hat \cO^{\Lambda_1 \m^{(1)} }_{i_1 j_1 } )^I_J  ( \hat
  \cO^\dagger{}^{{\L}_2 \m^{(2)}  }_{i_2 j_2 })^K_L}
& =\frac{ \delta^{ \m^{(1)} \m^{(2)} }  }{(n!)^2} \sum_{ \sigma_1 , \sigma_2 }  D^{ {\L}_1 }_{i_1j_1} ( \sigma_1 )  ~ 
 D^{ {\L}_2 }_{i_2j_2} ( \sigma_2 ) ~~  {\langle}  (\sigma_1 {\bf X}^{\m^{(1)}}    \sigma_1^{-1})^I_J 
(\sigma_2 { \bf  X}^{\m^{(2)}}   \sigma_2^{-1} )^{K}_L   {\rangle} \nn \\ 
& = \frac{ \delta^{ \m^{(1)} \m^{(2)} }  }{(n!)^2}  \sum_{ \sigma_1 , \sigma_2 } \sum_{ \gamma \in H_{\mu^{(1)} } } 
  D^{ {\L}_1 }_{i_1j_1} ( \sigma_1 )  ~ 
 D^{ {\L}_2 }_{i_2j_2} ( \sigma_2 )  ~~   ( \sigma_1   \gamma \sigma_2^{-1}  )^{I}_{L}
  ( \sigma_2 \gamma^{-1} \sigma_1^{-1} )_{J}^{K}   \nn \\ 
& =\frac{ \delta^{ \m^{(1)} \m^{(2)} } }{(n!)^2} \sum_{ \sigma_1 , \tau } \sum_{ \gamma \in H_{\mu^{(1)} } } 
  D^{ {\L}_1 }_{i_1j_1} ( \sigma_1 ) ~  
 D^{ {\L}_2 }_{i_2j_2} ( \tau^{-1} \sigma_1 \gamma  )  ~~   ( \tau )^{I}_{L} 
                       ( \tau^{-1}   )_{J}^{K}
\end{align}
The diagonality in $\mu^{(1)} ,\mu^{(2)} $ follows easily from Wick's theorem
and \eqref{basiccorrelatorUN}. 
In the last line we have replaced the $\sigma_2$ sum 
over $S_n$ by a $ \tau $ sum over $S_n$, where   $ \tau = \sigma_1 \gamma
\sigma_2^{-1}$. We can now  expand out $D^{ {\L}_2 }_{i_2j_2} ( \tau^{-1} \sigma_1 \gamma  )$
\begin{align}\label{covcor}  
\corr{ ( \hat \cO^{\Lambda_1 \m^{(1)} }_{i_1 j_1 } )^I_J  ( \hat
  \cO^\dagger{}^{{\L}_2 \m^{(2)}  }_{i_2 j_2 })^K_L }
& =\frac{ \delta^{ \m^{(1)} \m^{(2)} }  }{(n!)^2} \sum_{ \sigma_1 , \tau } \sum_{ \gamma \in H_{\mu^{(1)} } } 
  D^{ {\L}_1 }_{i_1j_1} ( \sigma_1 ) ~  
 D^{ {\L}_2 }_{i_2a} ( \tau^{-1}) D^{ {\L}_2 }_{ab} ( \sigma_1  ) D^{ {\L}_2 }_{bj_2} ( \gamma  )  ~  ( \tau )^{I}_{L} 
                       ( \tau^{-1}   )_{J}^{K} \nn \\
 & = \delta^{ \m^{(1)} \m^{(2)} }\delta^{{ \L}_1 {\L}_2 } ~\frac{|H_{\m^{(1)}}|}{n!d_{\L_1}} ~ D^{{\L}_1 }_{j_1 j_2
  } ( \Gamma  ) ~  \sum_{\tau } ~ D^{{\L}_1 }_{ i_1i_2 } ( \tau ) 
   ~  
  ( \tau )_{J}^{K}   ( \tau^{-1}   )^{I}_{L} 
\end{align} 
We have defined $\Gamma = \frac{1}{|H_{ \m^{(1)}} |}
\sum_{ \gamma \in H_{\mu^{(1)} } }
\gamma$, where $ |H_{\m^{(1)}}| = \mu^{(1)}_1 ! \cdots \mu^{(1)}_M ! $.  In the second
line we performed the sum over $\sigma_1 $ using identity
\eqref{schurorthogonalityorthogonal} from Appendix Section
\ref{sec:formulae} on formulae.

$D^{\L}_{jk}(\G)$ is a projector from the representation space of
$\Lambda$ onto the subspace which is invariant under 
$H_\m \equiv  H_{\mu^{(1)} } =  H_{\mu^{(2)}} $.
  The irreducible representation $ \Lambda $ of $ S_n$ gives, by
restriction, a representation of $H_\m$, generally reducible.  One can
decompose it in terms of irreps of $H_\m$. The projector $ \Gamma $
picks out the trivial irrep $ {\bf 1 } ( H_{\m} ) $ in this. This occurs with a
multiplicity $ g (\m ; \Lambda ) $. 
 We can write $D^{\L}_{j_1j_2}(\G) = {\langle} \L
, j_1 | \Gamma | \L , j_2 {\rangle} $ as 
\begin{equation}
 \label{gammaproj}  
 {\langle}   \L , j_1  |  \Gamma |    \L , j_2   {\rangle} 
 = \sum_{ \beta }  
{\langle} \L ,j_1|\Lambda ( S_n ) \rightarrow  { \bf 1 } (  H_\m )  ;  \beta   {\rangle} \;
{\langle} \Lambda ( S_n ) \rightarrow  { \bf 1 }( H_\m ), \beta  | \L ,
j_2 {\rangle} 
\end{equation}
The index $ \beta $ runs over an orthonormal basis for the multiplicity 
of the trivial irrep. ${ \bf 1 } (  H_{\m} ) $.  
Using the orthonormality of $\beta$ and inserting
  a complete set of states
\begin{align}
  \label{orthogbranch}   
\delta_{ \beta_1 \beta_2 } &=  {\langle} \Lambda ( S_n ) \rightarrow 
  { \bf 1 }(H_\m) , \beta_1  | \Lambda ( S_n )
 \rightarrow {\bf 1} (H_\m) , \beta_2  {\rangle}  \nn \\ 
& =\sum_{j=1}^{d_{\L} }  {\langle} \Lambda ( S_n ) \rightarrow {\bf 1}(H_\m) , \beta_1  | \Lambda , j {\rangle} \;  {\langle} \Lambda , j | \Lambda ( S_n ) \rightarrow {\bf{1}}(H_\m),
 \beta_2  {\rangle} 
\end{align}
This gives an orthogonality relation for the { \it branching
coefficients } $ {\langle} \Lambda ( S_n ) \rightarrow { \bf 1} (
H_\m) , \beta | \Lambda , j {\rangle} $. From the reality of the
symmetric group irreps.
\begin{align}\label{branchreality}  
 {\langle} \Lambda , j | \Lambda ( S_n ) \rightarrow { \bf 1 }(H_\m) , \beta {\rangle}
  =  {\langle} \Lambda ( S_n ) \rightarrow {\bf 1 } (H_\m) , \beta | \Lambda , j {\rangle} 
\end{align} 
See \cite{somers} and Appendix \ref{sec:branching} for calculations of
these branching coefficients.  To save space we shall define
\begin{equation}\label{savespace} 
  B_{j\b} \equiv {\langle} \Lambda , j | \Lambda ( S_n ) \rightarrow { \bf 1 }(H_\m)
  , \beta {\rangle} 
\end{equation}
It should be clear from the context which $\L$ and $\m$ are being
used.

Now we use these orthogonality properties to define
\begin{equation}\label{defofdiagonalcov}
\hat \cO^{\Lambda\m }_{i \beta} = \sum_{j } B_{j\b}
\;\hat \cO^{\Lambda\m}_{ij}
\end{equation}
Note that the number of operators $\hat \cO^{\Lambda\mu}_{i \beta } $ is
$ d_{\Lambda} g ( \mu ; \Lambda ) $ since $i$ runs over $d_{\Lambda }
$ values and $\beta $ runs over $ g ( \mu ; \Lambda ) $ values.  This
therefore correctly counts the covariant operators and gives the
explicit map \eqref{mapbasictofancy} from $V_M^{\otimes n }$ to a
state in $ V_{\Lambda}^{U(M)} \otimes V_{\Lambda} ^{S_n }$ for which
we have been looking,
\begin{align}\label{mapbasictofancyproper}
\hat \cO^{\Lambda\m }_{i \beta} = \sum_{j }\;B_{j\b}\; \frac{1}{n!}\sum_\s D^{ \L }_{ij} ( \sigma )
 \; \s\;{\bf X}^\m \; \s^{-1}
\end{align}
Note that together $\m$ and $\b$ give us the $U(M)$ state $m$.
Furthermore, using equations 
\eqref{covcor} \eqref{orthogbranch}\eqref{savespace} we see that 
 these operators $\cO^{\Lambda\m}_{i \beta}$ have a 
  simple 2-point function
\begin{align}\label{covcorbeta}  
\corr{ (  \hat  \cO^{\Lambda_1 \m^{(1)}  }_{i_1 \beta_1 } )^{I}_J    
(\hat \cO^\dagger{}^{\Lambda_2 \m^{(2)}  }_{i_2 \beta_2 } )^K_L } 
=  \delta^{ \m^{(1)} \m^{(2)} } \delta^{{ \L}_1 {\L}_2 }  \delta_{ \beta_1 \beta_2 } \frac{ |H_{\m^{(1)}}|}{n!d_{\L_1}} 
\sum_{\tau }    ~ D^{{\L}_1 }_{i_1 i_2 } ( \tau ) 
    ( \tau )_{J}^{K}   ( \tau^{-1}   )^{I}_{L}
\end{align} 

\subsection{Basis from projection}

We can also understand our basis as a projection onto a
linearly-independent subspace of the $\hat \cO^{\Lambda\m}_{ij}$.
The general theory is outlined in Appendix
\ref{sec:some-basic-linear}.

The set of operators $\hat \cO^{\Lambda\m}_{ij}$ appear to be giving
$d_\L^2$.  We know that the dimension of the representation $\L$ with
field content $\m$ is instead $d_\L g(\m;\L)$. The operators $\hat
\cO^{\Lambda\m}_{ij}$ are not in fact linearly independent.  This is
because permutations in $H_\m$ leave the $ { \bf X }^\mu $ fixed.
For any permutation $\gamma \in H_\m$, we have $ \cO^\m(\sigma )=
\cO^\m(\sigma \gamma)$. For the Fourier transform, we have by
relabelling $\sigma\rightarrow\sigma\gamma$ and using  
$D_{ij}^\Lambda(\sigma\gamma^{-1})=\sum_k
D_{ik}^\Lambda(\sigma)D_{jk}^\Lambda(\gamma)$, that 
\begin{align}
  \hat \cO^{\Lambda \m}_{ij}  =  D_{jk}^\Lambda(\gamma) \hat \cO^{\Lambda \m}_{ik}  \qquad \forall \gamma \in H_\m
\end{align}
Here the $j$ index belongs to  the representation $\Lambda$ and this
equation says that we seek the subspace which is invariant under the
subgroup $H_\m$.

So we seek operators, $\cO^{\Lambda \m}_{ij}$ such that
\begin{align}
  \hat \cO^{\Lambda \m}_{ij}  = \sum_k\hat \cO^{\Lambda \m}_{ik}
  D_{jk}^\Lambda(\Gamma)
\end{align}
which, using equations \eqref{gammaproj} \eqref{branchreality}
 \eqref{savespace},  becomes 
\begin{align}
  \hat \cO^{\Lambda \m}_{ij}  = \sum_{k,\beta}\hat \cO^{\Lambda \m}_{ik} B_{j\beta}B_{k\beta}
\label{Ointermsofbranching}
\end{align}

This is precisely the situation outlined
in~\ref{sec:some-basic-linear}. According to the analysis there, a
basis for such operators  is therefore given by
\begin{align} 
\hat \cO^{\Lambda \m }_{i \beta} = \sum_{j } B_{j\b}
\;\cO^{\Lambda \m}_{ij}
\end{align}
which is exactly the operator we found to have nice diagonality
properties for the two point function.

\subsection{Recovering the permutation basis}

Given the map \eqref{mapbasictofancyproper} from operators 
 $\hat{\cO}^\m(\s)$ of \eqref{exampleop}  labelled by permutations,  to
operators $\hat \cO^{\Lambda \m }_{i \beta}$ labelled by 
irreps $\L$ and states $i,j$, we would also like to invert  it,
i.e. to provide the map \eqref{mapfancytobasic} from $\hat
\cO^{\Lambda \m }_{i \beta}$ to
$\hat{\cO}^\m(\s)$. From~(\ref{Ointermsofbranching}) we have
\begin{align} \label{oob}
\hat \cO^{\Lambda \m}_{i
  j} =   \sum_\b  B_{j\b}
 \hat \cO^{\Lambda \m}_{i
  \beta}\ , 
\end{align}
so we simply have to find the inverse of the Fourier transform \eqref{PW} 
to give 
$\hat{\cO}^\m(\s)$ in terms of $\hat \cO^{\Lambda\m}_{i
  j}$
\begin{align}
  \hat{\cO}^\m(\s)=\sum_\Lambda \sum_{i,j} {d_\Lambda}
  D^\Lambda_{ij}(\sigma) \hat  \cO^{\Lambda \m}_{i
  j}  
\end{align}
This can be shown by using $\sum_{i,j}D_{ij}^\Lambda(\sigma_1)
D_{ij}^\Lambda(\sigma_2)=\chi_\Lambda(\sigma_1\sigma_2^{-1})$ and
that the expansion of the delta-function over the symmetric group  
in terms of characters is
\begin{align} 
\delta ( \sigma ) = \sum_{ \L } {d_{\L }\over n!} \chi_{ \L } ( \sigma ) \ .
\end{align}

Putting these together we find 
\begin{align} 
\hat{\cO}^\m(\s) = \sum_{\Lambda } \sum_{i,k } \sum_\b {d_{\Lambda }} D^{\Lambda}_{ik} ( \sigma ) B_{k\b}
 \hat \cO^{\Lambda \m}_{i
  \beta}\ .  \label{rewriteosm}
\end{align}
We know that $\hat{\cO}^\m(\s)$ span the vector space of all covariant
operators and they can be given as linear combinations of the
operators $\hat { \cO}^{\Lambda\m}_{i \beta}$ according
to~(\ref{rewriteosm}). Furthermore we know that the $\hat {
  \cO}^{\Lambda \m}_{i \beta}$ have the correct counting, and so we
have shown that they provide a basis for the space of all covariant
operators.

\subsection{$U(M) \times S_n $ transformations}

In this section we show that the index $i$ in $\hat \cO^{\Lambda
  \mu}_{i \beta } $ transforms according to irrep $\Lambda $ of the
symmetric group, and that the index $\beta$ (together with $\mu$)
transform according to the irrep $\Lambda $ of $U(M)$. For the reader
who is prepared to accept this, we recommend skipping ahead to Section
\ref{sec:invariant}.  To demonstrate this we need to show that the
$S_n$ action is a left-action on $\s$ and $U(M)$ is a right-action.

\subsubsection{$S_n$ action}

The action of $\tau \in S_n$ as defined in \eqref{eq:snactiononstateI}
is
\begin{equation}
  \hat \cO^\m(\s) \to \hat \cO^\m(\tau\s)
\end{equation}
Thus
\begin{align}
 \hat \cO_{ij}^{\L \m} & \to \sum_\s D_{ij}^\L(\s) \hat \cO^\m(\tau\s)  = \sum_\r D_{ij}^\L(\tau^{-1}\r )  \hat\cO^\m(\r) = D_{ik}^\L(\tau^{-1}) \hat\cO_{kj}^{\L \m}
\end{align}
It only acts on the $i$ index.

\subsubsection{$U(M)$ action}\label{sec:um-action}

In order to describe the action of $U(M)$ it is convenient 
to introduce the following operator, 

\begin{align} 
\hat \cO^{\Lambda}_{ij;a_1 \dots a_n}  = \sum_{\sigma } 
D^{ \L }_{ij} ( \sigma ) 
 \hat \cO_{\sigma;a_1 \dots a_n } 
\end{align}
where   $a_i$ are $U(M)$ indices and 
\begin{align}
  \hat  \cO_{\sigma;a_1 \dots a_n } = \sigma ~ \cdot ~  X_{a_1}
 \otimes X_{a_2} \cdots \otimes X_{a_n} ~ \cdot ~ 
 \sigma^{-1}   
\end{align} 
Under a $U(M)$ transformation $X_a \rightarrow \sum_{a'=1}^M g_{a}{}^{a'} X_{a'}$
and so 
\begin{align}
\hat \cO^{\Lambda}_{ij;\vec a} \rightarrow
  \sum_{\vec a} g_{\vec a}{}^{\vec a'} \hat \cO^{\Lambda}_{ij;\vec a'}
\end{align}
where $\vec a$ is short-hand for $a_1\dots a_n$ and $ g_{\vec
  a}{}^{\vec a'}$ stands for
$g_{a_1}{}^{a'_1}\dots g_{a_n}{}^{a'_n}$. This is the group action,
if instead we consider the
the action of the Lie algebra then we have $\delta
X_a=\sum_{a'=1}^M g_{a}{}^{a'} X_{a'}$ and $\delta \hat \cO^{\Lambda}_{ij;\vec a} =
  \sum_{\vec a} g_{\vec a}{}^{\vec a'} \hat \cO^{\Lambda}_{ij;\vec a'
  }\ , $, where $g_{\vec a}{}^{\vec a'}= \sum_{\alpha=1}^{n} \delta_{a_1}^{a'_1}\dots
 \delta_{a_{\alpha-1}}^{a'_{{\alpha}-1}} g_{a_\alpha}^{a'_\alpha}
 \delta_{a_{\alpha+1}}^{a'_{{\alpha}+1}}
\dots
\delta_{a_n}^{a'_n} \ ,$ but everything else follows.

As we saw in \eqref{exampleop} the operators $ \hat \cO ( \vec a ) $ 
can be written  as  $ \hat \cO^{ \mu } ( \sigma ) $.  
 The sum over all possible
values for $\vec a$ can be written as
\begin{align}
  \sum_{\vec a}  \hat \cO ({\vec a})=\sum_{\mu} \sum_{\sigma} 
{1\over |H_{\mu}|}
    \hat \cO^{\mu}  (\sigma )
\end{align}
where the division by $|H_{\mu}|$ accounts for the fact that
$ \vec a $ determines a unique  $\mu$ but not a unique  $\sigma$.

So under a $U(M)$ transformation we have 
\begin{align}
\hat \cO^{\Lambda  \mu }_{ij }  \rightarrow &\sum_{\vec a}
g_{\mu}{}^{\vec a'}  \hat \cO^{\Lambda}_{ij;{\vec a'}}\\
=& \sum_{\mu'}\sum_{\sigma'} {1\over |H_{\mu'}|} g_{\mu}{}^{\sigma'\mu'} \sum_{\sigma } D^{ \L }_{ij} ( \sigma )  
 \hat \cO^{\m^{\prime} }_{\sigma \sigma'  }\\
 =& \sum_{\mu'}\sum_{\sigma'} {1\over |H_{\mu'}|}
 g_{\mu}{}^{\sigma'\mu'} \sum_{\sigma } D^{ \L }_{ik} ( \sigma ) 
 \hat \cO^{\mu'}_{\sigma }\, D^{ \L }_{jk}(\sigma')\\
=& \sum_{\mu'}\sum_{\sigma'} {1\over |H_{\mu'}|} g_{\mu}{}^{\sigma'\mu'} D^{ \L }_{jk}(\sigma') 
\hat \cO^{ \L\mu' }_{ik } 
\end{align}
and we see that the $U(M)$ transformation leaves the
$\Lambda$ and $i$ unchanged.
The linearly independent operators
$\cO^{\Lambda\mu}_{i \beta }:=\sum_j \cO^{\Lambda\mu }_{ij} B_{j\beta}$
then transform as
\begin{align}
  \hat \cO^{\Lambda \mu}_{i\beta}\rightarrow&  \sum_{\mu'}\sum_{\sigma'}
  {1\over |H_{\mu'}|} g_{\mu}{}^{\sigma'\mu'}
\sum_j
  B_{j\beta} D^{ \L }_{jk}(\sigma') 
\hat \cO^{ \L \mu'}_{ik}\\
=& \sum_{\mu'}\sum_{\sigma'\in S_n} 
 \sum_{j,k} \sum_{\beta'} {1\over |H_{\mu'}|}
g_{\mu}{}^{\sigma'\mu'}  B_{j\beta} D^{ \L }_{jk}(\sigma')B_{k
    \beta'}  
\hat \cO^{ \L \mu'}_{i\beta'} 
\end{align}
Here to obtain the second line we have used~(\ref{oob}).

To summarise we can say that under a $U(M)$ transformation
\begin{align}
  \cO^{\Lambda \mu}_{i\beta}\rightarrow \sum_{\mu'} \sum_{\beta'}
  g_{\mu \beta}{}^{\mu'\beta'}  
\cO^{ \L \mu'}_{i\beta'} 
\end{align}
where
\begin{align}
  g_{\mu \beta}{}^{\mu'\beta'} =\sum_{\sigma'\in S_n} 
 \sum_{j,k} \sum_{\beta'} {1\over |H_{\mu'}|}
g_{\mu}{}^{\sigma'\mu'}  B_{j\beta} D^{ \L }_{jk}(\sigma')B_{k
    \beta'}
\end{align}
Again the Lie algebra follows  precisely from this by making $g$ infinitesimal.

\subsubsection{$U(M)$ highest weight  }

The highest weight state (HWS)  of the $U(M)$ representation has a
particularly simple form. The HWS has 
$\mu=\Lambda$ and since $g(\Lambda;\Lambda)=1$, $\beta$ takes only one
value. Therefore the projector $D_{ij}(\Gamma)$ projects onto a 
one-dimensional subspace. 
We can always choose representing matrices such that 
\begin{align}
  D_{ij}(\Gamma)=\delta_{1i}\delta_{1j}
\end{align}
giving the branching coefficients $B_{j\beta}=\delta_{1j}$.
Hence  the basis of highest weight states is given by the operators
\begin{align}\label{hws}
   \hat \cO^{\Lambda,HWS}_{i}:=\hat \cO^{\Lambda \Lambda}_{i \beta}=
   \hat  \cO^{\Lambda\Lambda}_{i1}=  \sum_\sigma D^\Lambda_{i1}(\sigma)    \s \; {\bf X}^\L \; \s^{-1}
\end{align}
and all operators are obtained by applying lowering operators to
these (the precise action of $U(M)$ on the operators is given in
Section~\ref{sec:um-action}).

It is common to give irreducible
representations of $U(M)$ by using Young projectors acting on the
tensor product of fundamental representations. This procedure is
closely related to our method - in particular the highest weight
state will be similar to~(\ref{hws}) with $D^\Lambda_{i1}$ replaced by
the Young projector $P_i^{\Lambda}$ -  but does not lead to a diagonal
two-point function.

\section{Gauge-invariant operators}\label{sec:invariant}

We will now construct the corresponding basis for gauge-invariant
operators made from multi-traces of multi-fields.  As before, it is
instructive first to consider the counting of such operators.

\subsection{Counting}

The formula for the number $N(\m_1, \dots \m_M)$ of gauge-invariants
operators (including multi-traces) made from fields $\mu_1$ of $X_1$,
$\mu_2$ of $X_2$, \dots $\m_M$ of $X_M$ at large $N$ is given via
P\'olya theory to be the coefficient of $x_1^{\m_1} \cdots x_M^{\m_M}$
in the partition function
\begin{equation}\label{eq:largeNpartition}
 \cZ_{U(N \to \infty)}(x_i) =   \prod_{k=1}^\infty \frac{1}{1 - (x_1^k + \cdots +
  x_M^k)} = \sum_\m N(\m_1, \dots \m_M) x_1^{\m_1} \cdots x_M^{\m_M}
\end{equation}
This coefficient is
\begin{equation}\label{eq:nicecounting}
  N(\m_1, \dots \m_M) = \sum_R \sum_{\L} C(R,R,\L)
  g(\m; \L)
\end{equation}
$R$ is a representation of $S_n$ and $\L$ is a representation of
$U(M)$ and $S_n$, where $n = \sum_{k} \m_k$.  $C(R,R,\L)$ is the
coefficient of $\L$ in the (inner) tensor product $R \otimes R$ of
$S_n$ representations.  In this formula the Littlewood Richardson
coefficient $g(\m; \L)$, see equation \eqref{eq:LRstate}, is counting
states in the $U(M)$ representation $\L$ with field content $\m$ and
$\sum_R C(R,R,\L)$ is counting the number of gauge-invariant
multiplets for $\L$.

This formula is discussed in \cite{Dolan:2007rq} for the finite $N$
case (developing the large $N$ results of \cite{sundborg,ammpr}),
where the Clebsch-multiplicities appear in properties of symmetric
polynomials, from the free field partition function. We give a short
proof here of the simpler large $N$ case.

$C(R,S,T)$ is given by
\begin{equation}
  C(R,S,T) = \sum_{C_{\bf i}\in S_n} \frac{1}{| \Sym (C_{{\bf i}})| }
  \chi_R(C_{{\bf i}}) \chi_S(C_{{\bf i}}) \chi_T(C_{{\bf i}})
\end{equation}
$C_{{\bf i}}$ represents the conjugacy class of $S_n$ with $i_1$
1-cycles, $i_2$ 2-cycles, \dots $i_n$ $n$-cycles.  $| \Sym (C_{{\bf
i}})|$ is the size of the symmetry group of the conjugacy class
$C_{{\bf i}}$.
We can immediately simplify $\sum_R C(R,R,\L)$ using the orthogonality
relation
\begin{equation}
  \sum_{R} \chi_R (\rho) \chi_R (\tau) = | \Sym(\tau) |
   \;\;\delta([\tau] = [\rho])
\end{equation}
to get
\begin{equation}
  \sum_R C(R,R,\L) = \sum_{C_{\bf i} \in S_n} \chi_{\L}(C_{{\bf i}})
\end{equation}

$N(\m_1, \dots \m_M)$ is the coefficient of $x_1^{\m_1} \cdots
x_M^{\m_M}$ in
\begin{align}
&  \prod_{k=1}^\infty \frac{1}{1 - (x_1^k + \cdots +
  x_M^k)} \\
  =& 
  \sum_{i_1, i_2, \dots} (x_1 + \cdots + x_M)^{i_1} \; (x_1^2 + \cdots +
  x_M^2)^{i_2}  \cdots 
\end{align}
For any symmetric polynomial $P$ the coefficient of $x_1^{\m_1}
\cdots x_M^{\m_M}$ is given by 
\begin{equation}
  [P]_\m = \sum_{\L} g(\m; \L) \left[\Delta \cdot P \right]_{[\L_1
  + M-1, \L_2 + M-2, \dots \L_M]}
\end{equation}
$\D$ is the discriminant $\D(x) = \prod_{i<j} (x_i - x_j)$.  This
formula is given in Chapter 4.3 of Fulton and
Harris\cite{fultonharris}.
Thus
\begin{align}\label{frob}
  N(\m_1, \dots \m_M) & =  \sum_{i_1, i_2, \dots} \sum_{\L}g(\m; \L) \left[\Delta \cdot
  \prod_j (x_1^j + \cdots x_M^j)^{i_j} \right]_{[\L_1
  + M-1, \L_2 + M-2, \dots \L_M]} \nn \\
  & = \sum_{i_1, i_2, \dots}  \sum_{\L} g(\m; \L)
   \chi_\L(C_{{\bf i}}) \nn \\
   & =  \sum_R \sum_{\L} C(R,R,\L)
  g(\m; \L)
\end{align}
In the second line we have identified the Frobenius character formula
\cite{fultonharris}. 
Note that the quantity summed here can also be recognised as the
character of the representation of $S_n$ induced from the trivial
representation of the subgroup $S_{\m_1} \times \cdots S_{\m_M}$
\begin{equation}
  \psi_{\L}(C_{{\bf i}}) =  \sum_{\L} g(\m; \L)
  \chi_\L(C_{{\bf i}})
\end{equation}
A more explicit proof is given in Appendix \ref{sec:explicit}.

\subsubsection{Finite $N$ counting}

Once we interpret $R$ as a representation of $U(N)$ the counting
formula
\begin{equation}
  N(\m_1, \dots \m_M) = \sum_R \sum_{\L} C(R,R,\L)
  g(\m; \L)
\end{equation}
extends to finite $N$, where we truncate the sum over $R$ to Young
diagrams with at most $N$ rows.  This formula was derived for the
finite $N$ case in \cite{Dolan:2007rq}.

If we decompose the partition function into $U(M)$ characters
\begin{align}
  \chi^\Lambda(x_i)= \sum_\mu g(\mu;\Lambda)x_1^{\mu_1} x_2^{\mu_2} \dots
  x_M^{\mu_M}
\end{align}
then we get
\begin{align}
   \cZ_{U(N)}(x_i)= \sum_R \sum_\Lambda C(R,R,\Lambda) \chi^\Lambda(x_i)
\end{align}
where the coefficient of each character $\chi^\Lambda$ in the
partition function gives the number of $U(M)$ multiplets $\Lambda$ in
the theory.

\subsection{Operators and correlators}

\subsubsection{The trace basis}

We start by defining multitraces $ \cO^{\m }(\a,\s)$ of the basic
gauge-covariant operator $\cO^\m(\s) $ where the permutation alpha
determines the cycles of the trace
\begin{equation}
 \cO^{\m  }(\a,\s) =  \tr(\a \hat \cO^\m(\s)) =  \tr(~ \a ~ \s\;
  X_1^{\m_1} \otimes  \cdots  \otimes X_M^{\m_M} \; \s^{-1} ) 
 = tr ( \alpha ~ \sigma ~ { \bf X }^{\mu} ~  \sigma^{-1} ) 
\end{equation}
The trace is being taken in $V^{ \otimes n } $.  
Using permutations $ \alpha $ to parametrize multi-traces
is useful and non-trivial already in the half-BPS case,
 where a single matrix is involved 
\cite{cjr}.  $\s \in S_n$ controls the order of the
fields and ties in closely with the $U(M)$ representation of the
operator.

This operator has two symmetries that leave it unchanged
\begin{equation} 
\begin{array}{rclcl}
(\a ,\sigma )& \rightarrow & (\pi \alpha \pi^{-1}, \pi \sigma)& & \pi \in S_n \nn \\
(\a ,\sigma )& \rightarrow &  (\a, \sigma \g)& & \g \in H_\m  = S_{\m_1} \times \cdots \times S_{\m_M} \subset S_n\label{tracesymmetries2}
\end{array}
\end{equation}
With the $\pi$ symmetry we can remove the $\s$ dependence of our
operator.  If we then take equivalence classes $[\a]$ of the
equivalence relation
\begin{equation}
  \a \sim \g \a \g^{-1} \label{eq:equivclass}
\end{equation}
for $\g \in H_\m$ then $\tr ( [\a]\; {\bf X}^\mu)$ is the independent
\emph{trace basis}.  It has the correct counting of gauge-invariant
operators for fixed $\m$ field content (see Section
\ref{sec:tracecounting}).

\begin{figure}[t]
\begin{center}
\includegraphics[trim=0 -20 0 -35]{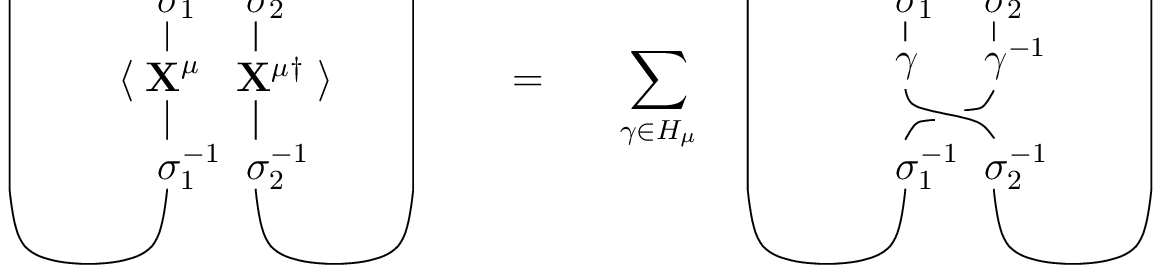}
\caption{two-point function for the gauge-invariant operator} \label{fig:twopointfunction}
\end{center}
\end{figure}

The two-point function of $\cO^{\m }(\a,\s)$ is given by
\begin{equation}
 \corr{ \cO^{\m^{(1) } } (\a_1,\s_1)\; \cO^\dagger{}^{ \m^{(2)} }
 (\a_2,\s_2)} 
 =  \delta^{\mu^{(1)}\mu^{(2)} }   \sum_{\g \in H_{\m^{(1)}}} \tr(\a_1 \s_1 \g
  \s_2^{-1} \a_2 \s_2 \g^{-1} \s_1^{-1}  )
\end{equation} 
This is easily derived from the diagrammatic presentation of
correlators as in Figure \ref{fig:twopointfunction}.  Such
diagrammatic manipulations are also useful in deriving properties of
half-BPS correlators \cite{cr}.

\subsubsection{A diagonal basis}\label{sec:ginvdiagbas} 

First we Fourier transform the $\s$ and multiply by the branching
coefficient as we did for the covariant case (from here on we will use
the Einstein summation convention)
\begin{equation}
  \cO_{i\b}^{\L \mu} (\a) =   \frac{1}{n!} \sum_{\s}B_{j \b}  D^{\L}_{ij} (\s) \tr(\a \s\;
  { \bf X }^{\mu}   \; \s^{-1} ) 
\end{equation}
$D^{\L}_{ij} (\s)$ is the orthogonal representing matrix.  This
operator is the trace with $\a$ of the covariant operator
\eqref{defofdiagonalcov}: $\cO_{i\b}^{\L\m}(\a) =
\tr(\a~\hat\cO_{i\b}^{\L\m})$.  The two point function follows from the
covariant result \eqref{covcorbeta}
\begin{equation}
 \corr{ \cO_{i_1\b_1}^{\L_1\mu^{(1)} }(\a_1)
  \cO_{i_2\b_2}^\dagger{}^{\L_2 \mu^{(2)} }(\a_2)
  } = \delta^{ \m^{(1)} \m^{(2)} } \delta^{{ \L}_1 {\L}_2 }  \delta_{ \beta_1 \beta_2 } 
 \frac{|H_\m| }{n!d_{\L_1}} 
 \sum_{\rho }    ~ D^{{\L}_1 }_{i_1 i_2 } ( \rho ) \tr(\a_1 \rho \a_2
  \rho^{-1} )
\end{equation}
The trace is being taken in $ V^{ \otimes n } $ and we have Schur-Weyl duality 
\bea\label{schweyUN}  
  V^{\otimes n } = \oplus_{ T } V_{ T }^{U(N)}  \otimes
 V_{ T } ^{S_n } 
\eea 
 So  we can use $\tr(
\s~ \mathbb{I}) = \sum_T \chi_T(\s) \chi_T(\mathbb{I}) = \sum_T
\chi_T (\s)\Dim T $, where $\Dim T $ is the dimension of the $U(N)$
representation $T $, to get
\begin{equation}
\delta^{ \m^{(1)} \m^{(2)} } \delta^{{ \L}_1 {\L}_2 }  \delta_{ \beta_1 \beta_2 }
  \frac{ |H_{\m^{(1)}}|}{n!d_{\L_1}} 
 \sum_{\rho }    ~ D^{{\L}_1 }_{i_1 i_2 } ( \rho )   \sum_T \Dim T ~
  D^T_{ab}(\a_1) D^T_{bc}(\rho) D^T_{cd}(\a_2) D^T_{da}(\rho^{-1})
\end{equation}
Now use the orthogonality of the matrices and identity
\eqref{sumsigmaid} on the $\rho$ sum
\begin{equation}
\delta^{ \m^{(1)} \m^{(2)} } \delta^{{ \L}_1 {\L}_2 }  \delta_{ \beta_1 \beta_2 }
  \frac{ |H_{\m^{(1)}}|}{d_{\L_1}^2} \sum_T \Dim T ~ D^T_{ab}(\a_1) D^T_{cd}(\a_2) \sum_{\tau} S^{\tau,}{}^{\L_1}_{i_1}\;{}^T_a\;{}^T_b  \;
 S^{\tau,}{}^{\L_1}_{i_2}\;{}^T_c\;{}^T_d \label{eq:introduceclebsch}
\end{equation}
The $S$ here are Clebsch-Gordan coefficients for the (inner) tensor
product of $S_n$.  Section \ref{sec:CGcoef} explains where these come
from in more detail.

Next we Fourier transform $\a$ in the same way to get the operator
\begin{equation}
  \cO_{i \b;}^{\L\mu}{}^R_{kl} =    \frac{1}{n!}\sum_{\a} D^{R}_{kl} (\a) \frac{1}{n!}\sum_{\s}B_{j \b} D^{\L}_{ij} (\s)
   \tr(\a \s\;
   { \bf X }^{\mu}  \; \s^{-1} ) 
\end{equation}
If we apply this sum over $\a_1$ in \eqref{eq:introduceclebsch} then
we can factor out
\begin{equation}
  \sum_{\a_1}  D^{R_1}_{k_1l_1} (\a_1) D^T_{ab}(\a_1) =
  \frac{n!}{d_{R_1}}\delta^{R_1T} \delta_{k_1 a} \delta_{l_1 b}
\end{equation}
We have used \eqref{schurorthogonalityorthogonal}. Thus we get
\begin{equation}
 \corr{ \cO_{i_1\b_1 ;}^{\L_1 \m^{(1)} }{}_{k_1l_1}^{R_1}
  \cO^\dagger{}_{i_2\b_2;}^{\L_2 \m^{(2)}  }{}_{ k_2l_2}^{R_2} }
  = \delta^{ \m^{(1)} \m^{(2)} } \delta^{{ \L}_1 {\L}_2 }  \delta_{ \beta_1 \beta_2 }\delta^{R_1 R_2 } 
  \frac{ |H_{\m^{(1)}}|  \Dim R_1}{d_{\L_1}^2d_{R_1}^2}  ~ \sum_{\tau} S^{\tau,}{}^{\L_1}_{i_1}\;{}^{R_1}_{k_1}\;{}^{R_1}_{l_1}  \;
 S^{\tau,}{}^{\L_1}_{i_2}\;{}^{R_1}_{k_2}\;{}^{R_1}_{l_2} \label{corijkl}
\end{equation}

Finally we define 
\begin{equation}\label{finform} 
  \cO^{ \L \mu , R }_{ \b, \tau } =  S^{\tau  ,}{}^{\L}_{ i }\;{}^{R}_{k}\;{}^{R}_{l}\;\;  \cO_{i\b;}^{\L \mu} {}^R_{kl}
\end{equation}
Note that $ \beta $ runs over $ 1 $ to $ g ( \m ; \Lambda ) $ and $
\tau $ runs over $1$ to $ C ( R , R ; \Lambda )$ which are the
factors appearing in the counting \eqref{eq:nicecounting}.  Some of
these operators are worked out in Appendix \ref{sec:exampleoperators}.

Using \eqref{corijkl} and the orthogonality of the Clebsch-Gordan
coefficients \eqref{ortho1}
\begin{align} 
 \corr{ \cO^{\L_1 \mu^{(1)} ,R_1}_{ \beta_1,\tau_1 } \cO^\dagger{}^{\L_2 \mu^{(2)}  ,R_2}_{ \beta_2 ,
  \tau_2 } } = 
\delta^{ \m^{(1)} \m^{(2)} }
\delta^{{\L}_1 {\L}_2 }  \delta_{ \beta_1 \beta_2 }\delta^{R_1 R_2 } \delta_{ \tau_1 \tau_2 }
  \frac{ |H_{\m^{(1)}}|  \Dim R_1}{d_{R_1}^2}\label{orthoresult} 
\end{align}

\subsection{Simplified operator}

The operator $\cO^{ \L\m , R }_{ \b, \tau }$ can be written more simply
in the trace basis, removing the $\s$ redundancy.  Take the operator
\begin{equation}
  \cO^{ \L \mu , R }_{ \b, \tau } =\frac{1}{(n!)^2} \sum_{\s,\a} B_{j \b} \; S^{\tau  ,}{}^{  \L }_{ i  }\;{}^{R}_{k}\;{}^{R}_{l}\;\; D_{ij}^\L(\s) D_{kl}^R(\a) \tr(\a\s\;
  {\bf X^{\mu} } \;\s^{-1} )
\end{equation}
and apply identity \eqref{Hamer186} to $ S^{\tau  ,}{}^{  \L }_{   i
  }\;{}^{R}_{k }\;{}^{R}_{l}\;\; D_{ij}^\L(\s)$ to get
\begin{align}
    \cO^{ \L \mu , R }_{ \b, \tau } & = \frac{1}{(n!)^2}\sum_{\s,\a} B_{j \b} \; S^{\tau  ,}{}^{  \L }_{  j
  }\;{}^{R}_{p }\;{}^{R}_{q}\;\; D_{kp}^R(\s)D_{lq}^R(\s) D_{kl}^R(\a) \tr(\a\s\;
  {\bf X}^{\mu} \;\s^{-1} ) \nn\\
& = \frac{1}{(n!)^2}\sum_{\s,\a} B_{j \b} \; S^{\tau  ,}{}^{  \L }_{  j
  }\;{}^{R}_{p }\;{}^{R}_{q}\;\; D_{pq}^R(\s^{-1}\a\s) \tr(\s^{-1}\a\s\;
  {\bf X}^{\mu}  )\label{simplifiedop-1}  
\end{align} 
In the second line we have exploited the fact that we are using
orthogonal representing matrices.  Finally we make a
substitution of summation variables and do the sum over $\s$
\begin{equation} 
   \cO^{ \L \mu , R }_{ \b, \tau }= \frac{1}{n!}\sum_{\a} B_{j \b} \; S^{\tau  ,}{}^{  \L }_{  j
  }\;{}^{R}_{p }\;{}^{R}_{q}\;\; D_{pq}^R(\a) \tr(\a\;
  {\bf X}^{\mu} )  \label{simplifiedop}
\end{equation}
It is easy to check that this operator is constant on the equivalence
classes of $\a$ given by \eqref{eq:equivclass}.

\subsection{Half-BPS case}

The Schur polynomial diagonalisation of \cite{cjr} is a special case
of this operator, when the representation is the trivial one $\L =
[n]$ and we consider the highest weight state of this representation
$\m_1 = n$.

In this case the counting is 
\begin{equation}
  N(n) = \sum_R \sum_{\L} C(R,R,\L)
  g([n]; \L) = \sum_R C(R,R,[n]) = \sum_R 1 = p(n)
\end{equation}
$p(n)$ is the number of partitions of $n$ and we have used the fact
that for the trivial representation $C(R,R,[n]) = 1$, since $R \otimes
[n] = R$.  Also $g([n];\L)$ is only non-zero for the trivial
representation $\L = [n]$, when it is one.

For the Clebsch-Gordan coefficient we have 
\begin{equation}
  S^{[n]}_{1}\;{}^{R}_{k}\;{}^{R}_{l}  = \frac{1}{\sqrt{d_R}}
  \delta_{kl}
\end{equation}
This follows from setting $T$ to be the trivial representation 
in \eqref{sumsigmaid} and then using the orthogonality relation
\eqref{schurorthogonalityorthogonal}. 
The operator is therefore 
\begin{equation}
  \cO^{ [n] , R } =  \frac{1}{\sqrt{d_R}}  \cO_{11}^\L{}^R_{kk} =  \frac{1}{\sqrt{d_R}} \chi_R(X_1)
\end{equation}

\subsection{Basis from projection}

All gauge invariant operators can be written as linear combinations of
operators of the form $\cO^\mu(\alpha)=\tr(\a X_1^{\otimes \m_1} \otimes  
\cdots \otimes 
 X_M^{\otimes \m_M} )$.  In order to obtain a basis of operators with the
correct counting we need to mod out by the symmetry
\begin{align}
  \cO^\mu(\alpha)=\cO^\mu(\gamma^{-1} \alpha \gamma)\qquad \forall
  \gamma\in H_\mu \label{sym}
\end{align}
Taking the Fourier transform we wish to find a basis for the space of operators
$\cO^{\m}{}^{ R}_{kl}:=\frac{1}{n!}\sum_\a D^R_{kl}(\a) \cO^\m(\alpha)$ up to the 
symmmetry
\begin{align}
  \cO^{\m}{}^{R}_{kl}=D^R_{km}(\gamma)D^R_{ln}(\gamma) \cO^{\m}{}^{ R}_{mn}\qquad \forall
  \gamma \in H_\m
\end{align}
which follows directly from~(\ref{sym}). 

Now $ \cO^{\m}{}^{R}_{kl}$ lies in the  tensor product of $S_n$ irreps. 
  $R\otimes R$
(carried by the two indices $k, l$)
and the problem becomes that of finding the subspace of this which is
invariant under $H_{\m}$. The projector which projects onto this space is
found by taking the sum of all the permutations in $H_{\m}$, so we are
seeking the set of operators $\cO^{\m}{}^{R}_{kl}$ for which
\begin{align}
 \cO^{\m}{}^{R}_{kl}&=P_{kl;mn} \cO^{\m}{}^{R}_{mn}\\
P_{kl;mn}&:={1\over |H_{\m} |} \sum_{\gamma\in H_{\m} }
D^R_{km}(\gamma)D^R_{ln}(\gamma) \label{projklmn1} 
\end{align}
If we write $D^R_{km}(\gamma)D^R_{ln}(\gamma)$ as a matrix in
$R\otimes R$ we can decompose it by inserting complete sets of states
\begin{align}
 {1\over |H_{\m} |} \sum_{\gamma\in H_{\m} } D^R_{km}(\gamma)D^R_{ln}(\gamma) & =
 \bra{R,k;R,l} \G \ket{R,m;R,n}  \nn \\
 & =  \sum_{\L,\tau; \L',\tau'}
 \braket{R,k;R,l}{\L,\tau,s}\bra{\L,\tau,s} \G\ket{\L',\tau',t}
 \braket{\L',\tau',t}{R,m;R,n} \nn \\
& =  \sum_{\L,\tau}
 \braket{R,k;R,l}{\L,\tau,s}\bra{\L,\tau,s} \G\ket{\L,\tau,t}
 \braket{\L,\tau,t}{R,m;R,n}
\end{align}
where $\Gamma=\sum_{\gamma \in H_\m} \gamma/|H_\m|$. The factors
$\braket{R,k;R,l}{\L,\tau,s}$ are just the Clebsch-Gordan
coefficients \eqref{clebschoverlap}, so we can write the projector
\begin{align}
  P_{kl;mn}=\sum_\Lambda \sum_\tau \sum_{s,t} D^\Lambda_{st}(\Gamma)
S^{\tau ,}{}^{\L}_{ s}\;{}^R_k\;{}^R_l S^{\tau ,}{}^{ \L}_{t}\;{}^R_m\;{}^R_n\label{projklmn2}
\end{align}
 We then write
$D^\L_{ts}(\Gamma)=B_{t\beta}B_{s\beta}$ as in~(\ref{gammaproj}) and so we can
decompose the projector into the form~(\ref{decomp})
\begin{align}
P_{kl;mn}=\sum_{\Lambda,\tau,\beta} b_{kl;\Lambda \tau \beta} b_{mn;\Lambda \tau \beta}
\end{align}
where
\begin{align}
  b_{kl;\Lambda \tau \beta}:=B_{s\beta} S^{\tau ,}{}^{ \L}_{s}\;{}^R_k\;{}^R_l
\end{align}
Now we have precisely the situation described in
appendix~\ref{sec:some-basic-linear}, with the relation~(\ref{bbd})
following from~(\ref{ortho1}). Therefore
\begin{align}\label{bso}
   \cO^{ \L\m , R }_{ \b, \tau } := b_{kl;\Lambda\tau\beta}\cO^{\m}{}^{R}_{kl}=B_{s\beta} S^{\tau ,}{}^{ \L}_{s}\;{}^R_k\;{}^R_l \cO^{\m}{}^{R}_{kl}
\end{align}
provides a basis of gauge invariant operators.

Notice that here we started by ignoring the the $U(M)$ aspect of the
operators completely and yet we find the $U(M)$ representation
$\Lambda$ appearing  as before.

\subsection{Recovering the trace basis}\label{sec:invrecovertrce}

It will be useful to invert the map so that we can write $
\cO^\m(\alpha)$ in terms of $\cO^{\L\m , R}_{ \beta, \tau } $.  Given that
$\cO^\m(\alpha) = \cO^\m(\g \alpha \g^{-1})$, for the operator
$\cO^{\m}{}^{R}_{kl}=\frac{1}{n!}\sum_\a D^R_{kl}(\a) \cO^\m(\alpha)$ we have
\begin{align}
  \cO^{\m}{}^{R}_{kl} & = \frac{1}{|H_\m|} \sum_{\g \in H_\m}
  D^R_{km}(\gamma)D^R_{ln}(\gamma) \cO^{\m}{}^{R}_{mn}  \nn \\
& = P_{kl;mn} \cO^{\m}{}^{R}_{mn} \nn \\
 & =\sum_{\Lambda,\tau,\beta} b_{kl;\Lambda \tau \beta} \cO^{ \L\m , R }_{ \b, \tau }
\end{align}
where we have used the projector identities in the previous section.

Now use
\begin{equation}
   \sum_R d_R D^R_{kl}(\a') D^R_{kl} (\a) = \sum_R d_R
   \chi_R(\a'\a^{-1}) = n! \delta(\s'\s^{-1})
\end{equation}
to get
\begin{equation}
  \cO^\m(\a) = \tr(\a\; {\bf X}^\m) = \sum_R d_R D^R_{kl} (\a) B_{s\beta} S^{\tau ,}{}^{ \L}_{s}\;{}^R_k\;{}^R_l \cO^{ \L\m , R }_{ \b, \tau }
\end{equation}
We know that the $\cO^\m(\a)$ span the vector space of all
gauge-invariant operators and they can be given as linear combinations
of the operators $\cO^{\Lambda\m,R}_{\beta,\tau}$. Since the
$\cO^{\Lambda\m,R}_{\beta, \tau}$ also have the correct counting, they
provide a basis.

\section{Including fermions}\label{sec:fermions}

It is interesting to extend these results from the case of Lie groups
$U(M)$ to super Lie groups $U(M_1|M_2)$. Indeed the space of eighth
BPS operators in $\cN$=4 SYM corresponds to the case $U(3|2)$,
the three scalar fields $X,Y,Z$ of the $U(3)$ sector combining with
two fermions, ${\bar \lambda}^1{}_2, {\bar \lambda}^1{}_1$ (in the
notation of\cite{Bianchi:2006ti}). The adjoint of the fermions $\bar
\l^1{}_{\dot a}$ is denoted $\l_{1a}$.
The two-point function of the two fermionic fields is then given by
\begin{equation}\label{basiccorrelatorfermionUN}
  \corr{ ({\bar \lambda}^1{}_{\dot a})^i_j (\lambda_{1a})^k_l} =
  \delta_{\dot aa}\, \delta^i_l \delta^k_j
\end{equation}
Note that here, as for the bosonic case,  we have ignored the $x$
dependence which is
 $(\delta_{a\dot a} x^0_{12}-\sigma^i_{a\dot a}x^i_{12} )/x_{12}^4$. By
taking a limit where  separation in time  $x^0_{12}$ dominates 
the separations in space $x^i_{12}$, we have that 
 the two-point function is proportional to $\delta_{a \dot a}$

The full set of fundamental fields in the sector is thus
denoted  $X_a$ as previously, but where $X_a$ is bosonic for $a=1\dots
M_1$ and fermionic for $a=M_1+1\dots M_1+M_2$. The main difference this makes
as far as we are concerned is that we pick up an extra minus sign when
two fermionic  fields are swapped. So
\begin{align}
(X_a)_i^j (X_b)_k^l= (-1)^{\epsilon(X_a)\epsilon(X_b)}(X_b)_k^l (X_a)_i^j 
\end{align}
where we have defined the  Grassmann parity of $X_a$ as
\begin{align}
&\epsilon(X_a)=0\qquad a=1\dots M_1 \\
&\epsilon(X_a)=1\qquad a=M_1+1\dots M_1+M_2
\end{align}

We will also find it useful to define the Grassmann parity of
permutations, given ${\bf X}^\mu$. We first define it for transpositions
\begin{align}
\epsilon((ij))&=0 \qquad i \ \rm{ or } \ j=1 \dots n_1\\  
\epsilon((ij))&=1 \qquad i \ \rm{ and } \ j=n_1+1 \dots n_1+n_2
\end{align}
and extend it to all permutations by insisting that 
\begin{align}
  \epsilon(\sigma\tau)=\epsilon(\sigma)+\epsilon(\tau) \quad \rm{ mod
  }\  2
\end{align}
Here $n_1=\sum_{k=1}^{M_1}\mu_k$ is the total number of bosonic
fields, and $n_2=\sum_{k=M_1+1}^{M_1+M_2}\mu_k$, the total number of
fermionic fields with $n=n_1+n_2$.  Then the covariant two point
function of operators $ {\bf X }^\m $
(cf. \eqref{eq:basicmultifieldcorr} for the bosonic case) contains the
following minus sign
\begin{align}
  {\langle}  ({\bf X }^\mu)^I_J 
({ \bf  X^\dagger }^\mu )^{K}_L   {\rangle}=(-1)^{\epsilon(\gamma)} ( \gamma
)^{I}_{L} 
  ( \gamma^{-1} )_{J}^{K} 
\end{align}
and this structure allows us to apply the results from the purely
bosonic case with only minor modification.

Firstly consider the gauge covariant operators. We will proceed in
precise analogy to the bosonic case.  
We may define operators precisely as in~(\ref{PW}) 
\begin{align} 
\hat \cO^{\Lambda \m}_{ij} = \frac{1}{n!}\sum_{\sigma }
D^{ \L }_{ij} 
( \sigma ) \hat \cO^\m(\sigma ) 
\end{align} 
 The manipulations of~(\ref{covcor}) then follow
through as previously but with the additional minus sign associated
with $\gamma$. This  additional minus sign accompanying $\gamma$ means
that in the final line of~(\ref{covcor}), the two-point function,  $\Gamma$ becomes
$\Gamma = \frac{1}{|H_\m|}\sum_{ \gamma \in H_{\mu} }
(-1)^{\epsilon(\gamma)}\gamma$. This means that $D^{\L}_{jk}(\G)$
becomes a projector from the representation space of
$\Lambda$ onto the subspace which is invariant under $H$ up to a
sign, ie
$D^{\L}_{ij}(\sigma)D^{\L}_{jk}(\G)=(-1)^{\epsilon(\sigma)}D^{\L}_{ik}(\G)$.
Since it is a projector this can be written in terms of branching
coefficients as in~(\ref{gammaproj}). The Kostka number becomes equal to the Littlewood-Richardson
coefficient for the appearance of
$\L$ in the tensor product of trivial single-row representations and
antisymmetric representations
$[\mu_1]\otimes \dots \otimes[\mu_{M_1}] \otimes 
[1^{\mu_{M_1+1}}] \otimes \dots \otimes [1^{\mu_{M_1+M_2}}]$. The formulae for the covariant
operators should all follow through also without additional changes as
they are essentially linear combinations of traces of the covariant  
operators already considered. 
In particular the operators defined in~(\ref{finform}) or
(\ref{simplifiedop}) are modified only in the afore-mentioned
modification of the branching coefficients.

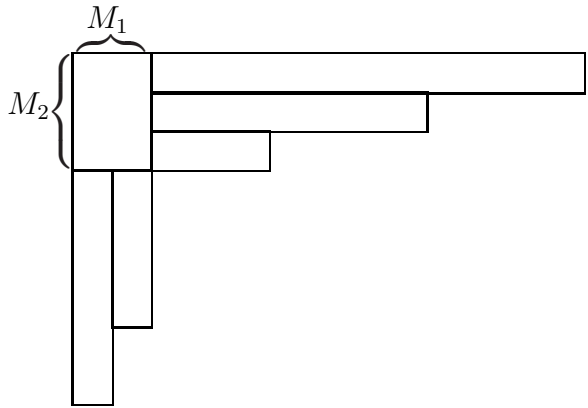
\begin{figure}
  \begin{center}
\setlength{\unitlength}{2144sp}%
\begingroup\makeatletter\ifx\SetFigFontNFSS\undefined%
\gdef\SetFigFontNFSS#1#2#3#4#5{%
  \reset@font\fontsize{#1}{#2pt}%
  \fontfamily{#3}\fontseries{#4}\fontshape{#5}%
  \selectfont}%
\fi\endgroup%
\begin{picture}(6327,4221)(1336,-5023)
\thinlines
{\color[rgb]{0,0,0}\put(2701,-1411){\framebox(4950,450){}}
}%
{\color[rgb]{0,0,0}\put(2701,-1861){\framebox(3150,450){}}
}%
{\color[rgb]{0,0,0}\put(2701,-2311){\framebox(1350,450){}}
}%
{\color[rgb]{0,0,0}\put(1801,-5011){\framebox(450,2700){}}
}%
{\color[rgb]{0,0,0}\put(2251,-4111){\framebox(450,1800){}}
}%
{\color[rgb]{0,0,0}\put(1801,-2311){\framebox(900,1350){}}
}%
\put(1051,-1661){\makebox(0,0)[lb]{\smash{{\SetFigFontNFSS{12}{14.4}{\rmdefault}{\mddefault}{\updefault}{\color[rgb]{0,0,0}$M_2$}%
}}}}
\put(1500,-1720){$\left\{
    \begin{tabular}{c}
      \ \\ \ \\ \ \\
    \end{tabular}
\right. $ }
\put(1850,-950){$\overbrace{\hspace{25pt}}$}
\put(1951,-661){\makebox(0,0)[lb]{\smash{{\SetFigFontNFSS{12}{14.4}{\rmdefault}{\mddefault}{\updefault}{\color[rgb]{0,0,0}$M_1$}%
}}}}
\end{picture}%
\end{center}
    \caption{Allowed shape for the Young tableau of the representations $\Lambda$ of $U(M_1|M_2)$.}\label{fig:1}
\end{figure}

Furthermore the counting formula will be identical 
to~(\ref{frob}) namely
\begin{align}
  N(\m_1, \dots \m_{M_1+M_2})
    =  \sum_R \sum_{\L} C(R,R,\L)
  g(\m; \L)
\end{align}
the only difference being  in  the definition of the Kostka number and
in the allowed representations $\Lambda$.
The allowed $U(M_1|M_2)$
representations  $\Lambda$ have the shape as shown in
Figure~\ref{fig:1}. 
The first $M_1$ rows are unbounded, but rather
more unusually, the first
$M_2$ columns are also unbounded. See, for example~\cite{Bars:1983se}
for more information on representations of supergroups and supertableaux. 

\subsection{Single fermion}

The simplest example involving fermions is given by
$U(M_1|M_2)=U(0|1)$ corresponding to a single fermion. The allowed
representations $\Lambda$ are the totally antisymmetric reps,
$\Lambda=[1^n]$ and the counting becomes
\begin{equation}
  N(n) = \sum_R \sum_{\L} C(R,R,\L)
  g([1^n]; \L) = \sum_R C(R,R,[1^n])=\sum_{R=\tilde R}1
\end{equation}
where the final sum indicates a sum over self-conjugate
representations. This follows from the fact that $R \otimes [1^n] =
\tilde R$, where $\tilde R$ is the partition conjugate to $R$,
obtained by exchanging the rows and columns of $R$.

One can count the allowed operators for $N>n$ as follows. Single trace
operators must have an odd number of fields (otherwise they vanish,
for example $\tr(\psi \psi)=\psi_i^j \psi_j^i=-\psi_j^i\psi_i^j=0$).
Multitrace operators are then made of single-trace operators with an
odd number of fields in each, with the restriction that you cannot
have the same single trace term twice (otherwise it vanishes by
anti-symmetry). So all our operators have the form
\begin{align}
  \tr(\psi^{2k_1+1})  \tr(\psi^{2k_2+1})\dots   \tr(\psi^{2k_l+1}) \qquad
k_1>k_2>\dots >k_l\geq 0
\end{align}
The map between these operators and self-conjugate Young-tableaux with
$k_j+j$ boxes in the $j$th row and column gives a one-to-one
correspondence between multi-trace operators of a single matrix-valued
fermion and self-conjugate Young tableaux.

\section{Extremal higher-point  correlators }\label{extcorr} 

\begin{figure}[t]
\begin{center}
\resizebox{!}{6cm}{
\includegraphics{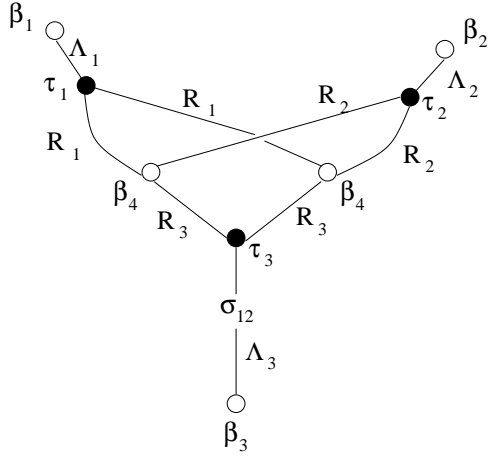} }
\caption{Diagram for three-point function} \label{fig:three-point-ext}
\end{center}
\end{figure}

We have, using the simplified form of the operators in (\ref{simplifiedop})
\bea\label{3ptbranch}  
&&  \langle \cO^{\L_1 \m^{(1)}  , R_1 }_{\beta_1 , \tau_1 } (x_1) ~ 
 \cO^{\L_2 \m^{(2)} , R_2 }_{  \beta_2 , \tau_2 } (x_2)  ~
 \cO^{\dagger ~ \L_3 \m^{(3)} , R_3}_{ \beta_3 , \tau_3 } (0)  \rangle  \cr 
&& = 
  x_1^{-2n_1} x_2^{-2n_2 }  \delta^{ ( \mu^{(1)} +\mu^{(2)} ) , \mu^{(3)} }
 { | H_\m| \Dim R_3 \over d_{R_1} d_{R_2} d_{R_3}  }
  \cr 
&& 
    ~ S^{\tau_1 ,  \L_1~  R_1 ~ R_1}_{~~~j_1 ~~  p_1 ~  q_1 } 
 ~S^{\tau_2 , \L_2 ~ R_2 ~  R_2}_{~~~j_2 ~~ p_2 ~ q_2 } 
 ~ S^{\tau_3,  \L_3 ~ R_3 ~ R_3}_{~~~j_4 ~~ p_3 ~ q_3 } \cr 
&& D^{\L_3 }_{j_4 j_3} ( \sigma_{12} )~  B_{j_3 \beta_3 }
 B_{j_1 \beta_1 } B_{j_2 \beta_2 }  ~  B^{R_3 \rightarrow R_1\circ R_2 ; \beta_{4}  }_{~q_3; ~~p_1~ p_2 } 
   ~ B^{  R_3 \rightarrow R_1\circ R_2 ; \beta_4  }_{~p_3;~~ q_1~ q_2 }
\eea 
All repeated indices in the RHS, which do not appear on the LHS, are
summed.  This is conveniently represented in a diagram as in Figure
\ref{fig:three-point-ext}.  The diagram contains two kinds of
vertices. Filled circles labelled by $ \tau_1,\tau_2,\tau_3 $
represent inner-product couplings (or Clebsch-Gordan coefficients).
Open circles represent outer-product couplings (or branching
coefficients) and are labelled by $ \beta_1 , \beta_2 .. $.  The
diagram gives a simpler (with fewer indices) and accurate
representation of (the non-trivial part in the last two lines of)
 formula (\ref{3ptbranch}) from which the formula
can be reconstructed unambiguously.  To reconstruct the formula from
the diagram, we insert indices for states along the edges of the
diagram.  We write $S$-factors coupling the indices incident on a
filled circle and $B$-factors coupling the indices incident on the
open circles. When an open circle has a single incident line, it
stands for a branching of a symmetric group irrep into the trivial
irrep of a specified sub-group determined by the $\mu $ vector
associated with the corresponding operator. We could make the $\mu $
dependence explicit in these vertices of the diagram or in the
equivalent $B_{j \beta }$ in the formula, but we have followed the
convention of the earlier parts of the paper, of leaving that $\mu $
dependence implicit. The permutation $ \sigma_{12} $ re-orders the
sequence of $X,Y,Z $ coming from the two operators $\cO^{\L_1\m^{(1)}
,  R_1}_{\beta_1, \tau_1} $ and $\cO^{\L_2\m^{(1)}, R_2}_{\beta_2, \tau_2} $. They
lead to operators $\tr_{V^{\otimes ( n_1+n_2 ) }} ( ( \alpha_1 \cdot
\alpha_2 ) {\bf X}^{\mu^{(1)}} \cdot {\bf X}^{\mu^{(2)}} ) $ which can be
re-written as $\tr_{V^{\otimes ( n_1+n_2 ) }} ( ( \a_1 \cdot \a_2 )  \sigma_{12}
{\bf X}^{\mu^{(1)}+ \mu^{(2) }} \sigma_{12}^{-1} ) $.

The product of branching coefficients is obtained from 
the sum over matrix elements.
\bea 
&& { d_{R_1} d_{R_2}  \over n_1 !  n_2 ! }  \sum_{ \alpha_1 , \alpha_2 } 
D^{R_1}_{p_1q_1}  ( \alpha_1 )  D^{R_2}_{p_2q_2} ( \alpha_1 )
D^{R_3}_{p_3q_3} ( \alpha_1 \circ \alpha_2  ) \cr 
&& = \sum_{\beta_4 } 
  B^{R_3 \rightarrow R_1\circ R_2 ; \beta_{4}  }_{~q_3; ~~p_1~ p_2 } 
   ~ B^{  R_3 \rightarrow R_1\circ R_2 ; \beta_4  }_{~p_3;~~ q_1~ q_2 }
\eea 
The branching coefficients express the change of basis from that of
states in an irrep $R_3$ of $S_{n_3} $ to that of states in an irrep $
R_1\circ R_2 $ of the $ S_{n_1} \times S_{n_2} $ subgroup of $ S_{n_3}
$ (where $n_3 = n_1 + n_2 $).

\begin{figure}[t]
\begin{center}
\resizebox{!}{8cm}{
\includegraphics{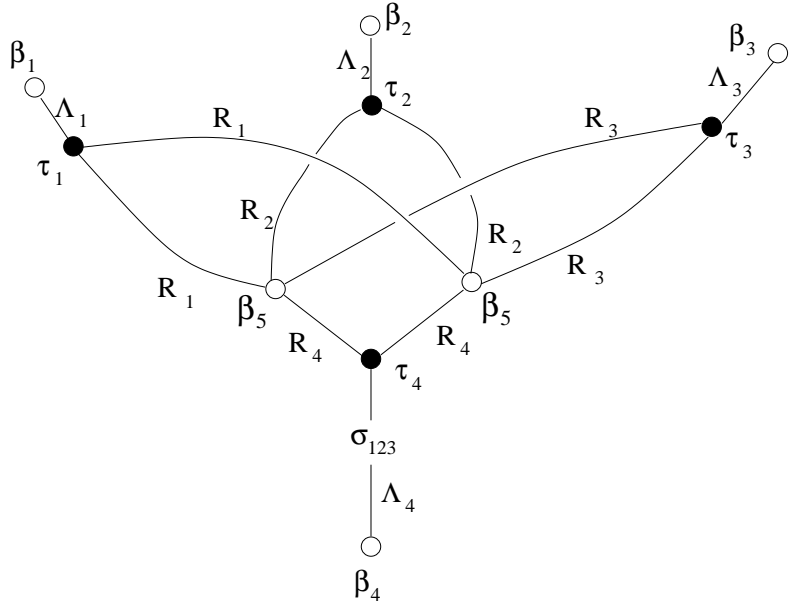} }
\caption{Diagram for four-point function} \label{fig:four-point-ext}
\end{center}
\end{figure}

\begin{figure}
\begin{center}
\resizebox{!}{8cm}{
\includegraphics{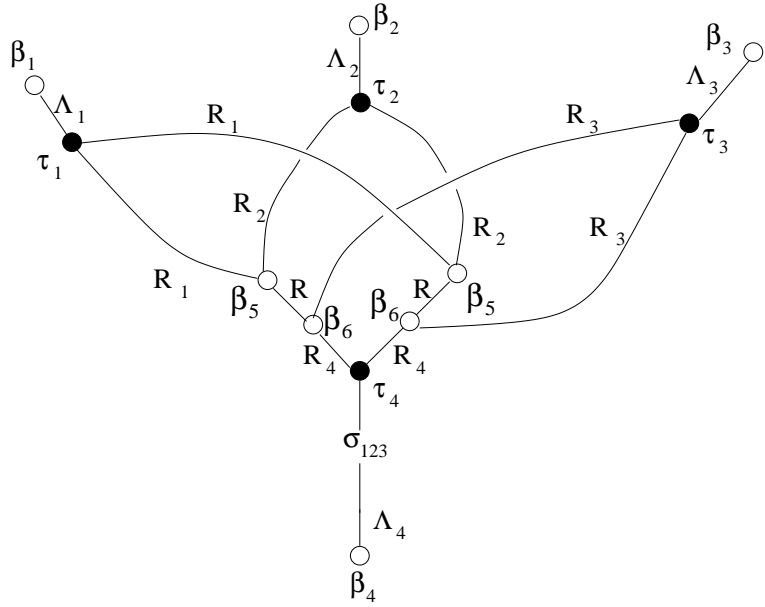} }
\caption{Splitting the 4-valent vertex into trivalents}
 \label{fig:4ptsplitvert}
\end{center}
\end{figure}

The extremal 4-point function can be written in terms of branching
coefficients generalising those of (\ref{3ptbranch}), now involving
4-valent branching vertices of irreps of $S_{n_1 +n_2 + n_3 }  $ into $
S_{n_1} \times S_{n_2 } \times S_{n_3} $.  The diagram is an easily
guessed generalisation of Figure \ref{fig:three-point-ext} which is
given as Figure \ref{fig:four-point-ext}. 
Here $ \sigma_{123}$ is a permutation which re-orders 
$ { \bf X }^{ \m^{(1)} } \cdot   { \bf X }^{ \m^{(2)} }
 \cdot  { \bf X }^{ \m^{(3)} } $ into 
$  { \bf X }^{ \m^{(1)} +  \m^{(2)} + \m^{(3)} } =  { \bf X }^{ \m^{(4)} } $.
  By performing  the
reduction of $S_n$ into $ S_{n_1 + n_2 } \times S_{n_3} $ and then the
reduction of $ S_{n_1 + n_2 } $ to $ S_{n_1 } \times S_{n_2} $ we can
split the 4-valent vertices into pairs as in Figure
\ref{fig:4ptsplitvert}.

\begin{figure}
\begin{center}
\resizebox{!}{6cm}{
\includegraphics{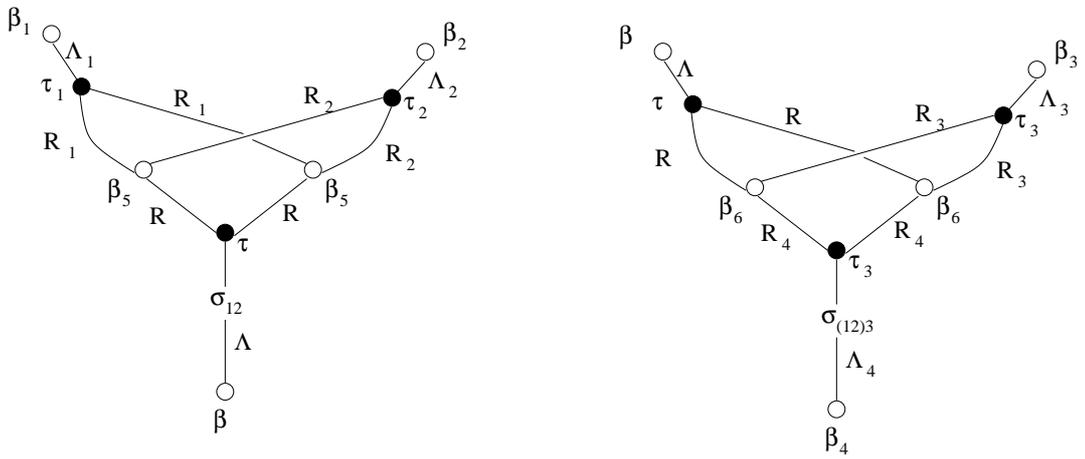} }
\caption{4-point as product of 3-points}\label{fig:433}
\end{center}
\end{figure}

Another approach to the 4-point function is to express 
the product of the first two operators 
 in terms of 3-point functions as follows
\bea\label{3ptcoeff}  
&&   \cO^{\L_1 \m^{(1)}, R_1}_{ \beta_1 , \tau_1 } (x )  
 \cO^{\L_2 \m^{(2)}  , R_2 }_{\beta_2 , \tau_2 } (x )   \cr 
&& =  \delta^{\m^{(1)} + \m^{(2)} , \m } \sum_{ R , \L , \beta , \tau } 
{    \langle \cO^{\L_1 \m^{(1)} , R_1 }_{\beta_1 , \tau_1 } (x )   
 \cO^{\L_2 \m^{(2)}  , R_2 }_{\beta_2 , \tau_2 } (x )  
  \cO^{\dagger ~ \L \mu ,  R }_{\beta , \tau }(0)   \rangle    
\over 
 \langle \cO^{ ~ \L \m , R}_{ \beta , \tau } ( 0 ) 
 \cO^{\dagger ~ \L \m ,R }_{ \beta , \tau } ( x ) \rangle }
~ \cO^{\L \m , R}_{ \beta , \tau } ( x )
\eea 
When we write out the LHS of (\ref{3ptcoeff}) we have 
a product of traces $ \tr_{V^{\otimes n_1} } ( \alpha_1 {\bf X}^{ \mu^{(1)} } ) 
 \tr_{ V^{\otimes n_2 }} ( \alpha_2 {\bf X}^{  \mu^{(2) } }  )  $.
 This can be re-written and manipulated as follows 
\bea 
&& \tr_{V^{\otimes n_1 + n_2 } } (  ( \a_1 \cdot \a_2 )  \sigma_{12} {\bf X}^{ \mu^{(1)} +   \mu^{(2)} } \sigma_{12}^{-1} ) \cr 
&& = \sum_{R_3}  { d_{R_3} \chi_{R_3} ( \alpha_3 ) \over n_3 ! } 
      \tr_{ V^{\otimes n_3 } } (   \alpha_3  ( \a_1 \cdot \a_2 ) \sigma_{12} {\bf X}^{ \mu^{(3)} } 
\sigma_{12}^{-1} ) \cr 
&& = { 1 \over | H | } \sum_{\gamma } \sum_{R_3}  { d_{R_3} \chi_{R_3} ( \alpha_3 ) \over n_3 ! } 
     \tr_{ V^{\otimes n_3 } } (  \alpha_3  ( \a_1 \cdot \a_2 )  \sigma_{12} \gamma {\bf X}^{ \mu^{(3)} } \gamma^{-1}  \sigma_{12}^{-1} ) 
\eea 
In the second line we have defined $n_3 = n_1 + n_2 $ , 
$ \mu^{(3)} =  \mu^{(1)} +   \mu^{(2)} $ and inserted a complete set of 
projectors for Young diagrams in $V^{\otimes n_3}$. In the third line 
we have used the symmetry of $ {\bf X}^{\mu^{(3)}} $ 
 under conjugation by elements in $H_3$ 
 to insert a group average. After a redefinition of 
$  \gamma^{-1}  \sigma_{12}^{-1}  \alpha_3  ( \a_1 \cdot \a_2 )  \sigma_{12} \gamma  = \tilde \alpha_3 $ it is easy to recognise the equality (\ref{3ptcoeff}).

If we derive the 4-point extremal correlator by using (\ref{3ptcoeff})
twice, we get a sum over $ \L , R , \beta , \tau $ of Figure
\ref{fig:433}. This is also equivalent to Figure
\ref{fig:four-point-ext} and Figure \ref{fig:4ptsplitvert}.  All these
equivalences can be seen as simple moves on the diagrams. This is
reminiscent of applications of group-theoretic quantities to topology
via topological field theories, where such moves are related to
topological invariances of triangulations and such
\cite{turvir,cfs,reshtur}.  In this case the precise topological
meaning of these correlators remains to be clarified.  Perhaps
relating the correlators to a topological field theory - which would have to
be one based on symmetric groups (constructed along the lines of
\cite{dijkwit}) -  might be a way to progress in this direction.  The
relation to one-dimensional matrix models described in Section
\ref{sec:redmat} may also be helpful in uncovering the topological meaning.

\section{Finite $N$ projector on traces and dual bases }\label{sec:finiteNproj}

We have found  that the complete set of holomorphic 
gauge invariant operators can be described in terms of 
a trace basis  $ \tr ( \alpha \;  { \bf X^{\m}  }    ) $ or 
in terms of a  basis $ \cO^{  \L \mu , R }_{\beta , \tau } $ which diagonalises 
the two-point function.  
In this section we will use  the index $B$ for the set 
$ ( \L , R , \beta , \tau ) $. The content $ \mu $ and the corresponding 
symmetry group $H_\m$ are understood to be fixed. 
We will use the index $A$ for equivalence classes 
of $ \alpha $ under  conjugation by permutations $ \gamma \in H_\m $.  
So we are using  the notation $\cO_{A} $ for the
traces and $ \cO_{B}$ for the orthogonal basis. We have 
\bea 
&&\cO_{B} = \sum_{A} F_{BA} \cO_{A} \cr 
&& \cO_{A} =  \sum_B G_{AB } \cO_{B} 
\eea  
where 
\bea\label{FG}  
&& F_{BA } =   { D^R_{pq} (\alpha ) \over n! } ~ B_{j \beta } ~ 
S^{\tau,}{}^{\L }_{j}\;{}^R_p\;{}^R_q \cr 
&&G_{AB} =  d_R  { D^R_{pq} (\alpha ) } ~ B_{j \beta } ~ 
  S^{\tau,}{}^{\L }_{j}\;{}^R_p\;{}^R_q
\eea 
As a side remark, note that if we rescale $ \cO^{\L\mu,R}_{\beta, \tau } $ 
to have a  $ \sqrt{ d_R \over n ! } $ instead of $ { 1 \over n! } $ 
then $F_{BA  } = G_{AB }$.   
We  can show, from \eqref{FG}   that 
\bea 
&& \sum_{A } F_{B_1 A} G_{AB_2}   =  \delta ( B_1 , B_2  )
\eea 
Indeed 
\begin{align}
& \sum_{\alpha } { d_R \over n!}  D^{R_1}_{p_1q_1} ( \alpha ) B_{j_1\beta_1} 
    S^{\tau_1,}{}^{\L_1 }_{j_1}\;{}^{R_1}_{p_1}\;{}^{R_1}_{q_1}
     D^{R_2}_{p_2q_2} ( \alpha ) B_{j_2\beta_2} 
    S^{\tau_2,}{}^{\L_2 }_{j_2}\;{}^{R_2}_{p_2}\;{}^{R_2}_{q_2} \cr 
& = \delta^{R_1R_2}  
S^{\tau_1,}{}^{\L_1 }_{j_1}\;{}^{R_1}_{p_1}\;{}^{R_1}_{q_1}
S^{\tau_2,}{}^{\L_2 }_{j_2}\;{}^{R_1}_{p_1}\;{}^{R_1}_{q_1}
B_{j_1 \b_1} B_{j_2 \b_2 }  \delta_{j_1 j_2 }   \cr 
&= \delta^{R_1 R_2} \delta_{\tau_1 \tau_2 } 
\delta^{\L_1 \L_2} \delta_{\beta_1 \beta_2 } \cr 
&= \delta ( B_1 , B_2  ) 
\end{align}
We used the orthogonality relations 
(\ref{schurorthogonalityorthogonal}), (\ref{ortho1}),  \eqref{orthogbranch}. 
Consider now 
\bea 
&& \sum_{B} G_{A_1B} F_{BA_2 } 
=  \sum_{ R , \L , \beta , \tau  }  { d_R \over n!}  D^{R_1}_{p_1q_1} ( \alpha ) B_{j_1\beta_1} 
    S^{\tau_1,}{}^{\L_1 }_{j_1}\;{}^{R_1}_{p_1}\;{}^{R_1}_{q_1}
     D^{R_2}_{p_2q_2} ( \alpha ) B_{j_2\beta_2} 
    S^{\tau_2,}{}^{\L_2 }_{j_2}\;{}^{R_2}_{p_2}\;{}^{R_2}_{q_2} \cr 
&&= \sum_{ R , \L ,  \tau  }  { d_R \over n!}  D^{R_1}_{p_1q_1} ( \alpha_1 ) 
        D^{R_2}_{p_2q_2} ( \alpha_1 )  
 D^{\L}_{j_1j_2} ( \Gamma ) 
   S^{\tau_1,}{}^{\L_1 }_{j_1}\;{}^{R_1}_{p_1}\;{}^{R_1}_{q_1} 
   S^{\tau_2,}{}^{\L_2 }_{j_2}\;{}^{R_2}_{p_2}\;{}^{R_2}_{q_2} \cr 
&& =  \sum_{ R }\sum_{\gamma } { d_R \over |H| }  D^R_{p_1q_1} ( \alpha_1 ) 
 D^R_{p_2q_2} ( \alpha_2 ) D^R_{p_1p_2} ( \gamma ) 
 D^R_{q_1q_2} ( \gamma ) \cr 
&& =  { 1 \over  |H | }\sum_{\g}  \sum_{ R } { d_R \over n ! }  
\chi_R ( \alpha_1 \gamma \alpha_2 \gamma^{-1} ) 
\eea 
We first recognised  the product of branching coefficients 
as the matrix element of a projector as in (\ref{gammaproj}). 
Then we used the identity of the expressions in 
(\ref{projklmn1}) and (\ref{projklmn2}). 
When $n < N $, the sum over $R$ runs over all  irreps 
of $S_n$ and we  have 
\bea 
  \sum_{ B  } G_{A_1 B } F_{BA_2} &&= 
{ 1 \over |H|  } \sum_{\gamma} 
 \delta ( \alpha_1 \gamma \alpha_2 \gamma^{-1} ) \cr 
&& = \delta ( A_1 ,  A_2  )  \qquad \qquad  \hbox { for large N } 
\eea  
When the large $N$ condition  is violated, i.e. when $ n > N $, 
the sum over representations only runs over Young diagrams 
with first column $c_1( R ) $ obeying $c_1 ( R ) \leq N$,
 which does not exhaust the 
irreps of $S_n$. In  this case we check that 
\bea\label{projfN}  
 P_{A_1A_2} \equiv  { 1 \over  |H | } \sum_{\g } \sum_{R : c_1 ( R ) \le N }  
 { d_R \over n ! }  \chi_R ( \alpha_1 \gamma \alpha_2 \gamma^{-1} ) 
\eea 
is a symmetric projector  
\bea 
 P_{A_1A_2} &=& P_{A_2 A_1 } \cr 
 \sum_{A_3} P_{A_1A_2 } P_{A_2A_3} &=& P_{A_1 A_3} 
\eea 
To summarise 
\bea 
\sum_{B} G_{A_1 B} F_{BA_2} = P_{A_1A_2} 
\eea 
where $P$ is the identity at large $N$, but is 
a projector at finite $N$.
This projector has the property that it leaves the trace basis 
invariant 
\bea\label{fNtrcons}  
\sum_{A_2}P_{A_1 A_2 } \cO_{A_2 } = \cO_{A_1 } 
\eea
This can be obtained directly  by using the Schur-Weyl 
duality relation \eqref{schweyUN} and recalling that 
for $U(N)$ the Young diagrams are restricted to have 
first column of length less than or equal to $N$.   
\bea 
 \tr ( \alpha { \bf X  }^{\mu}   )& = &
\sum_{ R: c_1(R) \le N } \tr  ( P_{ R } \alpha { \bf X }^{\mu}  )   \cr 
& =& \sum_{ R: c_1(R) \le N } \sum_{\alpha_1} { d_R \over n! } 
\chi_R ( \alpha_1 )  \tr  ( \alpha_1 \alpha { \bf X }^{\mu}  ) \cr 
&  =& \sum_{ R: c_1(R) \le N }  { d_R \over n! } 
 \chi_R ( \alpha_1 \alpha^{-1} ) 
     \tr ( \alpha_1  { \bf X }^{\mu}  ) \cr 
& =& {1 \over |H_{\mu} | }   \sum_{ R: c_1(R) \le N } \sum_{ \alpha , \gamma } 
    { d_R \over n! }    \chi_R ( \gamma \alpha_1 \gamma^{-1} 
 \alpha^{-1} )  \tr ( \alpha_1  { \bf X }^{\mu}  ) 
\eea 
In the last line we used the invariance of 
$ \tr ( \alpha_1  { \bf X }^{\mu}  )$ 
under conjugation by $ \gamma $. To go to the form in (\ref{fNtrcons}) 
simply recall that we are using $A$ 
for the $H_{\mu}$-equivalence classes of $\alpha $.  A similar discussion 
of finite $N$ projectors appears in the context of 2D Yang Mills 
in \cite{baeztaylor}. 

It is instructive to define a 
\emph{dual basis} to the traces \cite{brown,copto}
\bea\label{dualbasis}  
\cO_{A}^* = \sum_{ B }
 { F_{BA  } \over \langle B B^{\dagger} \rangle }  \cO_{B} 
 = \sum_{B , A_1 } { F_{BA  } \over \langle B B^{\dagger} \rangle }
  F_{B A_1} \ \cO_{A_1} 
\eea 
It follows that 
\bea 
 \langle \cO_{A_1}^*  \cO_{A_2}^{\dagger }  \rangle 
&=& \sum_{B_1 , B_2 }  
{ F_{ B_1 A_1  } G_{A_2 B_2 }  \over  \langle B_1 B_1^{\dagger} \rangle }
\langle \cO_{B_1} \cO^\dagger_{B_2} \rangle  \cr 
&=& P_{A_1 A_2 } 
\eea

\section{Applications to gauge-gravity duality }\label{sec:physics}

\subsection{Chiral ring;  N particles in 3D SHO } 

Consider the space of holomorphic operators contructed from 
$ X, Y , Z $. We have  given a basis labelled by  
$ \cO^{ \L\m , R }_{\beta , \tau } $. In this basis finite $N$ 
effects are simply encoded by the  the condition that the Young diagram 
$R$ has no more than $N$ rows. By the operator-state correspondence 
of CFT, these holomorphic gauge invariant operators map to a 
vector space $V$ of states. Now set to zero the commutators. 
\bea 
[ X , Y ] = [ Y , Z ] = [ X , Z ] = 0 
\eea 
A subspace  of the space of operators will vanish. 
This gives a subspace of $V$. Let us call this subspace $U$.
Given a subspace, we can define a quotient space $V/U$ 
defined as the set of equivalence classes of vectors 
modulo  addition of any vector in $U$. We have an exact 
sequence 
\begin{equation} 
0 \rightarrow U \rightarrow V \rightarrow V/U \rightarrow 0 
\end{equation}
Note that we do not have a natural map from $V/U$ to $V$.  $V/U$ is
not a subspace of $V$. The quotient space corresponds by the
operator-state relation to the chiral ring: operators modulo the
equivalence of setting commutators to zero.  There is no natural map
from $V/U$ to $V$ because there is no fixed choice for the
representatives of the equivalence class, so that for the chiral ring
we could choose say the ordered trace, e.g. $\tr(XYYZ)$, or the
symmetrised trace, e.g. $\Str(XYYZ)$.

By choosing some representatives of the equivalence classes, we can
define a map from $V/U$ to $V$. This map is not uniquely defined.  But
this won't matter. Having chosen the representatives we have an
isomorphism between $ V $ and $ U \oplus V/U$.  There is a basis
$\tilde A = ( \tilde A_s , \tilde A_a ) $ where $s$ runs over the
equivalence classes, and $a$ over a basis for the null-space. The
operators $ \cO_{A_s } $ are chosen representatives of the chiral
ring. The operators $ \cO_{A_a } $ span the operators which vanish
when commutators are set to zero.

Let us now restrict our considerations to $ n< N $.  Let $\cO_A$ be
any convenient basis, say the trace basis.  It can be enumerated using
P\'olya theory. For small $n$ it is easy to write the trace basis by
hand. Since we know the transformation between the trace basis and the
orthogonal basis $\cO_{B}$ explicitly, we have duals of the trace
basis according to \eqref{dualbasis}.

The $ \cO_{ \tilde A } $ are related to $ \cO_{ A } $ 
by some invertible transformation 
\bea 
\cO_{\tilde A } = \sum_A S_{\tilde A  A } \cO_{ A } 
\eea 
The inverse $T_{A \tilde A } $ satisfies 
\bea 
&& \sum_{A }  S_{\tilde A A } T_{A \tilde A^{\prime}  } = \delta_{ \tilde A  
\tilde A^{\prime}  } \cr 
&& \sum_{ \tA } T_{ A \tilde A } S_{ \tilde A A^{\prime}  } = \delta_{ A \prime A } 
\eea 
The dual basis for the $\cO_{\tilde A } $ is given by 
\bea
\cO_{\tilde  A }^* = \sum_{ A }  T_{A \tilde A }  \cO_{A}^* 
\eea 
After using  the expression \eqref{dualbasis} for $ \cO_{A}^* $ we have  
\bea\label{dualbas} 
\cO_{\tilde  A }^* = 
\sum_{B ,A, A_1 } { F_{BA  } T_{A \tA }   F_{B A_1} 
 \over \langle B B^{\dagger} \rangle }
 \ \cO_{A_1} 
\eea 
Indeed we check 
\bea\label{tAtAstardiag}  
  \langle \cO_{\tilde  A }^*   \cO^\dagger_{\tilde A^{\prime}  } \rangle 
&& = \sum_{A , A^{\prime}  }
  T_{A \tilde A }     S_{\tilde A^{\prime}  A^{\prime}  }
 \langle  \cO^*_{A} \cO^\dagger_{A^{\prime} }  \rangle  \cr 
&& = \sum_{ A  } 
 S_{\tilde A^{\prime}  A}  T_{A \tilde A  }  = \delta_{  \tilde A^{\prime} \tilde A } 
\eea

 For $n> N $, the $H_{\m} $-equivalence classes of traces form an 
 overcomplete set of operators. They have to be projected 
 using the finite $N$ projector \eqref{projfN} 
 in Section \ref{sec:finiteNproj}. 
 Nevertheless the outcome of the above discussion \eqref{dualbas} 
can  be reproduced in the finite $N$ case. 
 As before, after choosing representatives of the chiral 
ring, we have a basis of operators $ \cO_{\tilde A } $, which 
includes those vanishing when commutators are set to zero, labelled 
 by $\tilde A_{a} $,   
along with the representatives of chiral ring $ \cO_{ \tilde A_s }$. 
There is an invertible transformation which relates this basis 
to the orthogonal basis 
\bea  
\cO_{\tilde A } &=& \sum_{B } D_{\tA B } \cO_{B} \label{tAtoB}\\  
\cO_{B } &=& \sum_{ \tA } D^{-1}_{B \tA } \cO_{\tA }  \label{BtotA}
\eea 
The sum over $B$ is constrained by $ c_1 ( R ) \leq N $. 
A dual basis for the $ \cO_{\tilde A } $ can be given 
as 
\bea\label{dualtotA}  
\cO_{\tA}^* = \sum_B D^{-1}_{B \tA } {  \cO_{B} \over \langle B B^\dagger \rangle }   
\eea 
It is easy to check that 
\bea\label{dualtilde}  
\langle \cO_{\tA }^* \cO^\dagger_{ \tA'}  \rangle 
= \delta_{ \tA \tA'} 
\eea 
Now we would like to express the duals  in \eqref{dualtotA} 
in terms of a transformation of the redundant, non-orthogonal, yet 
convenient trace basis $ \cO_A$.  Expanding  $ \cO_B $ in terms of $\cO_{A} $ 
in \eqref{tAtoB} and using  \eqref{BtotA} in the equation 
for $\cO_A$ in terms of    $ \cO_B $, we get  
\bea 
 \cO_{\tA  } &=&   \sum_{ A } S_{\tA A }\cO_{A} \cr  
 \cO_{A } &=& \sum_{\tA} T_{A \tA} \cO_{\tA} 
\eea 
where $S, T$ are defined by 
\bea\label{STDFG}  
S_{\tA A } &=& \sum_B D_{\tA B } F_{B A } \cr 
T_{A \tA } &=& \sum_B G_{AB} D^{-1}_{B\tA } 
\eea 
Unlike the large $N$ case, $S,T$ are not inverses of each other, 
but instead we can derive 
\bea 
&& \sum_{ A } S_{\tA_1 A } T_{A \tA_2 } = \delta_{ \tA_1 \tA_2 } \cr 
&& \sum_{ \tA } T_{A_1 \tA } S_{\tA A_2 }    = P_{A_1 A_2 } 
\eea 
where $ P_{A_1 A_2 } $ is the finite $N$  projector defined in 
Section \ref{sec:finiteNproj}.  Using \eqref{STDFG} we can obtain an expression 
for $D^{-1} $ as 
\bea 
D^{-1}_{B \tA } =   \sum_{ A} F_{B A } T_{A \tA } 
\eea
This allows us to rewrite \eqref{dualtotA} as 
\bea\label{dualbasN} 
\cO_{\tA}^* = \sum_{B,A_1 , A }  { F_{B A_1 } T_{A_1 \tA } F_{B A }\over 
 \langle B B^\dagger \rangle  } \cO_{A} 
\eea 
This is identical in form to \eqref{dualbas} but now understood 
in a finite $N$ context.

The duality equation \eqref{dualtilde}   
can be unpacked to give 
\bea\label{listeq}  
&& \langle \cO_{\tA_s}^* \cO_{\tA_a}^{\dagger} \rangle  =  0 \cr 
&&  \langle \cO_{\tA_{s_i} }^* \cO_{\tA_{s_j } }^{\dagger} \rangle  =  
\delta_{ij}  \cr 
&&  \langle \cO_{\tA_{a_i} }^* \cO_{\tA_{a_j } }^{\dagger} \rangle  =  
\delta_{ij} 
\eea 
From the first equation it is clear that the space of operators spanned 
by $ \cO_{A_{s_i} }^*$ gives the orthogonal complement of the null space 
 $U$. This orthogonal complement is unique. It does not depend 
on the choice of representatives of the chiral ring. Note that we have 
characterised the orthogonal complement without having to diagonalise 
the metric on the space of states spanning $U$. 
We do not expect this orthogonality relation to be renormalised and
indeed this has been shown at 1 loop for certain quarter BPS operators
in~\cite{D'Hoker:2003vf}. The 
operators containing  commutators (corresponding to the subspace $U$) are the
descendants of long operators at all values of the coupling (they will
mix amongst themselves but the space $U$ remains invariant). 
Therefore since we do not expect mixing between short operators and
long operators, the 
space orthogonal to $U$ will also not change. 

It is very interesting  to consider the metric 
\bea\label{metshort}  
\langle \cO_{\tA_{s_i}}^* \cO_{\tA_{s_j}}^{\dagger \ast} \rangle  
\equiv  G^*_{ij } 
\eea 
on the short operators.
We have 
\bea\label{elabmet}  
 \langle \cO_{\tA}^* \cO_{\tA^{\prime} }^{\dagger \ast} \rangle &=& \sum_{B }
{  D^{-1}_{\tA B }  D^{-1}_{\tA^{\prime} B } \over \langle B B^{\dagger} \rangle } \cr 
&=& \sum_{ B, A_1 , A_2  }
 { F_{B A_1 } F_{B A_2 } \over \langle B B^{\dagger} \rangle } 
        T_{A_1 \tA} T_{A_2 \tA^{\prime}} 
\eea 
When $ \tA , \tA^{\prime}$ are restricted to the  $ \tA_{s_i},\tA_{s_j}$ 
i.e. to the subspace $V/U$, then \eqref{elabmet} gives $G^*$. 
 The metric  \eqref{metshort} 
 should not be renormalised. These non-renormalisation
theorems have been shown at one
loop by direct calculation in the case of quarter BPS operators of
dimension up to eight~\cite{Ryzhov:2001bp,D'Hoker:2003vf}. They can be proved using the
insertion formula, first applied to the correlation 
functions of  half-BPS in~\cite{Intriligator:1998ig,Eden:1999gh} and
applied to more general protected 
operators such as the quarter- and eighth-BPS operators considered here
in~\cite{3pnt,sym}. Furthermore it was shown there that the
 extremal higher-point correlators should also
be non-renormalised. It would be very interesting however to have some
checks of these  non-renormalisation theorems by direct calculation at higher
loops and for more general operators.
The other equations in (\ref{listeq})
 will receive corrections at higher orders in $g_{YM}^2 $.
Further progress on the metric \eqref{metshort} 
can be obtained by finding simplifications of the quantity 
\bea 
\sum_{B } {  F_{B A_1 } F_{B A_2 } \over \langle B B^{\dagger} \rangle }
\eea 
appearing in \eqref{elabmet}. 

For large $N$, i.e.  when the dimensions of the operators $n$ is less
than $N$,   we can choose as representatives of the chiral ring, the
products symmetrised traces, or we could choose products of ordered
traces $\tr ( X^{*} Y^{*} Z^{*}) $.  Either choice will lead to the
same (unique) orthogonal complement of the null-space by the above
construction.  For finite $N $, the essential part of the above story
goes through, but it is worth spelling out where the additional
subtleties lie.  There is still a well-defined subspace $U$
corresponding to vanishing operators when commutators are set to zero.
There is a well-defined quotient space $V/U$. By choosing
representatives of the chiral ring, we have an isomorphism from $V$ to
$ V \oplus V/U $. In this case we are not committing ourselves to
choosing symmetrised traces as representatives, since there could be
complicated relations among them at finite $N$.  All we need is that
there exists some choice of representatives. The form of possible
convenient choices will follow by looking more explicitly at the form
of the finite $N$ projector \eqref{projfN}. 

The quarter and eighth-BPS gauge invariant operators should be related 
to giant gravitons generalizing the analogous connection in the half-BPS 
case. It has been argued that 
the physics  of the eighth-BPS giants  \cite{mik}
 is given by the dynamics of $N$ particles 
in a 3D simple harmonic oscillator \cite{beasley}\cite{bglm}\cite{mansur}. 
States of harmonic oscillator system 
\bea\label{shostates}  
\prod_{i=1}^{N } a^{i  ~\dagger }_{n_{i1} , n_{i2} , n_{i3} } | 0 \rangle
\eea 
The index $i$ labels the particles. The natural numbers 
$(n_{i1} , n_{i2} , n_{i3} )$ label the  excitations 
along the $x, y , z $ direction for the $i$'th particle.
When we take an overlap of such a state  with excitations 
$n_{ia} $ with the conjugate of another state with excitations 
$n^{\prime}_{ia} $ we get an answer proportional to
\bea\label{niorth}  
 \prod_{ia} \delta  ( {n_{ia} } ,  {n^{\prime}_{ia} } )
\eea
In the leading large $N $ (planar) limit there is a simple 
map between the harmonic oscillator states and gauge invariant operators, 
which preserves the metric. 
The above SHO states can be  associated with 
\bea 
\prod_{i=1} \Str ( X^{n_{i1 }} Y^{n_{i2} } Z^{n_{i3} }  )| 0 \rangle 
\eea 
In the leading large $N$ (planar) limit, it does not actually matter whether we
choose symmetrised traces or ordered traces. This because different
trace structures do not mix, and mixings between different orderings
within a trace are also subleading in $1/N$.  With either choice, we
have the orthogonality \eqref{niorth} following from correlators of
gauge invariant operators.  But this does not work at subleading
orders in $1/N$ or at finite $N$.

The discussion above, in terms of
duals, gives a formula \eqref{elabmet} 
 for the  metric $G^*$ for the short representations.
This formula works to all orders in the $1/N$ expansion 
or at finite $N$.  
The formulae given above in terms of 
$F_{BA} , G_{AB} , S_{ \tA A } , T_{A \tA} $  give a way to 
work out $G^*$. After one diagonalises $G^*$, there should be a  
map from the orthogonal basis of short operators  to the SHO states 
\eqref{shostates}. The correct explicit form of such a  map remains 
an interesting open problem. We have  only offered  a 
recipe based on the $ F,G , S , T $ which can be applied in a case by
 case basis to write down $G^*$.

We also expect that 
extremal correlators 
\bea 
\langle \cO_{A_{s_{1}} }^{*} ( x_1 )  \cO_{A_{s_{2}} }^{*} ( x_2 ) 
 \cdots  \cO_{A_{s_{n}} }^{*} ( x_n ) 
\cO_{A_{s_{n+1 }} }^{\dagger *  } ( x_{n+1} ) \rangle 
\eea
are non-renormalised. They are expected to receive no corrections away
from their zero-coupling values, either at 1-loop or higher loops or
non-perturbatively. At strong coupling they are related to
supersymmetric eighth-BPS giant gravitons.  The correct map between
linear combinations of operators $ \cO_{A_{s} }^{*} $ and observables
in the N-particles in 3D-SHO should also allow a matching of these
extremal correlators.

A better understanding of $G^*$ should also 
help in making contact with quarter and eighth-BPS
generalisations of LLM geometries. These have been discussed 
recently in \cite{chenetal,gmno}. For example the geometric description 
of boundary conditions required for regularity of solutions, given 
in \cite{chenetal} should have a counterpart 
in the parametrisation of corresponding 
gauge invariant operators with metric \eqref{metshort}.  
Combining the  results on three-point functions in \ref{extcorr} 
along with the projection to the chiral ring in \ref{sec:finiteNproj}  
should lead to the three-point functions relevant to the strong coupling
limit. Following the arguments developed in \cite{skentay} 
these three-point functions contain detailed information 
about the generalised LLM geometries.

\subsection{Supergravity and $ \cO^{\L \mu , R}_{\beta , \tau } $ } 

In the above discussion we have discussed the spacetime 
interpretation of  the zero-coupling gauge-invariant operators 
by first considering their weak-coupling counterparts, which 
involves a projection to the orthogonal subspace to 
the operators which vanish when commutators are set to zero. 
A direct space-time interpretation of  the zero-coupling 
diagonal basis  $ \cO^{\L \mu , R}_{\beta , \tau } $ 
would be desirable. Of course by AdS/CFT duality 
such a description must exist, but because $ R=  ( g_s N)^{1/4} l_s $ 
this description will be at strong curvature, where SUGRA methods 
are not applicable.  In this tensionless limit of string theory there
is also an interesting supersymmetric version of the Higgs effect
discussed in ~\cite{sunborg,sezgin,bianchi,bouffe}, which relates to
some of the holomorphic operators we have considered becoming part of
long representations.

We may consider the S-duality $g_s \rightarrow { 1 \over g_s } $ which
will map us to the large radius regime. In usual discussions of
AdS/CFT, one uses the S-duality to fix $g_s < 1 $. Any $R$ is then
accessible by tuning $N$. However, when the objects of interest
include finite $N$ effects as a function of coupling as well as a
transition from zero to infinitesimal coupling, it is natural to
consider the large $g_s$ region at finite $N$. By the S-duality the
jump in SUSY states from zero to infinitesimal coupling will then
become a jump from finite radius to the flat space limit. It would be
interesting to explore supergravity approaches to this regime, and
perhaps in this context the zero-coupling diagonalisation, and the
associated parameters $ ( \L ,\m , R , \beta , \tau ) $ can be
interpreted in terms of parameters of spacetime solutions.

\subsection{Reduced multi-matrix model }\label{sec:redmat}

There is a reduced complex multi-matrix model obtained 
by reducing  the free action for the scalars of 4D $N=4 $ 
SYM on $S^3 \times R $. 
\bea 
\int dt ~\sum_a  \tr  ~ \partial_t  X_a   \partial_t  X_a^{ \dagger}  +  \tr  X_a  X_a^{ \dagger}
\eea 
The Gauss Law constraint requires a projection to gauge invariant 
(traced) states. 
Following \cite{cjr} (see also \cite{taktsu,djr,emergent}),   the Hamiltonian 
is 
\bea 
H = \sum_a  \tr (   A^{\dagger}_{a} A_{a}  +  B^{\dagger}_{a}   B_{a}  )
\eea 
The index $a$ is a flavour index which runs from $1$ to $3$. 
The operators $A_a , B_a$ also carry matrix indices and obey 
\bea 
&& [ A_{a}{}_j^{i} , A_{b}^\dagger{}_l^{ k  }] 
= \delta_{ab} \delta^i_l  \delta^k_j \cr 
&& [ B_{a}{}_j^{i} , B_{b}^\dagger{}_l^{ k  }] 
= \delta_{ab} \delta^i_l  \delta^k_j \cr 
&& [ A , A ] = [ B , B ]  = 0 
\eea 
The BPS states satisfy $ E = J_1 + J_2 + J_3 $. 
The $U(1)\times U(1) \times U(1) $ charges are 
\bea 
J_a = \tr (   A^{\dagger}_{a} A_{a}  -    B^{\dagger}_{a} B_{a} )
 \eea 
The BPS states are constructed by acting with 
$A^{\dagger} $ only, and no $ B^{\dagger}$. 
We can construct operators in this matrix quantum mechanics 
by replacing the $ {\bf X }^\m$ of earlier sections with sequences 
built from $ A^{\dagger} $. The $ \mu_1 $ copies of $ X_1 $, $ \mu_2 $  of  
$X_2$ etc are replaced by $ \mu_1$ of $A^{\dagger}_{1} $ etc. 
Our results on the diagonality of correlators imply immediately 
that 
\bea 
\langle 0 | \cO^{\L_1\m^{(1)} , R_1 }_{\beta_1 , \tau_1 }  ( A ) 
 \cO^{\L_2\m^{(2)} , R_2 }_{\beta_2 , \tau_2 } ( A^{\dagger} )  | 0  \rangle 
=\delta^{\m^{(1)}\m^{(2)}} \delta^{ \L_1 \L_2 } \delta^{R_1 R_2 } \delta_{\beta_1 \beta_2 }
 \delta_{\tau_1 \tau_2 } 
\eea 
Similarly our results on extremal correlators have counterparts 
in this multi-matrix quantum mechanics.

From the matrix oscillators we can construct composites 
$E_{ab} , T^i_j $ which obey $U(N)$ and $U(M)$ Lie algebra 
relations under commutation. 
\bea 
&& E_{ab} = A_a^{\dagger}{}_{j}^{i} A_b{}^j_{i} +  B^\dagger_{a}{}_{j}^{i} B_b{}^j_{i}
 \cr 
&& T^i_j =  A_a^{\dagger}{}_{k}^{i} A_a{}^k_{j} +  B^\dagger_{a}{}_{k}^{i} B_a{}^k_{j}
\eea 
Higher Hamiltonians can be constructed 
as $ \tr ( T^n ) $ and $ \tr ( E^m )$ which commute with the 
Hamiltonian $H$. Their eigenvalues, when acting 
on states, 
$  \cO^{\L\m , R }_{\beta, \tau } ( A^{\dagger} )  | 0  \rangle $ 
are invariants  (Casimirs) depending on $R , \L $ respectively. 
It is an interesting question whether there are conserved charges 
which measure $\tau $ and $\beta $ directly. Such conserved charges 
may help in finding a spacetime interpretation following the 
discussion of \cite{bcls} in the half-BPS case. These multi-matrix models
have also been discussed in \cite{agpol} where they were 
related to Calogero models. Expressing our diagonal operators in 
the fermionic variables may be useful in clarifying the 
integrable structure.

\section{Conclusions and open questions }

We have given a diagonal basis for correlators
in an $N \times  N  $ matrix theory with $U(M)$
global  symmetry. This has applications in the zero coupling 
limit of $\cN=4$ SYM. Holomorphic operators in the three complex 
spacetime-scalar matrix fields are  BPS at zero coupling.
For the two-point functions of these holomorphic operators 
with  their conjugates, we have a diagonal basis. 
 Some of these operators become part of 
long representations at weak coupling. We have given a characterisation of 
the orthogonal space to these long operators, which are genuine 
eighth-BPS operators. Our discussion is 
 valid at finite $N$. This gives a  framework which can be used 
for  the comparison of computations in the N-particle  3D-SHO, 
which comes from eigenvalues dynamics, with the gauge invariant operators. 
In particular extremal correlators of the genuine eighth-BPS 
operators computed from the gauge-invariant set-up of this paper 
should agree with those computed from eigenvalue dynamics. 
Explicit formulae for the two-point functions of these genuine-BPS 
operators can be obtained from this paper, but a natural diagonalisation 
on this subspace remains an open problem. This will be useful in setting up 
a comparison with results from eigenvalue dynamics. It should also 
allow a better understanding of the 
mapping from the geometry of giant graviton moduli spaces 
to the space of diagonal gauge-invariant eighth-BPS operators at finite $N$, 
generalising the geometrical explanation \cite{mst} of the stringy exclusion 
principle \cite{malstrom}.

It would be desirable to have a spacetime interpretation for the
diagonal basis in the full set of holomorphic operators. By AdS/CFT
there should certainly be a stringy interpretation. Whether there is
an interpretation in the supergravity limit is a very interesting
question. We have speculated, using S-duality that the answer is
positive. Progress on these issues would be fascinating because it
would give a spacetime interpretation to the parameter $ \tau $ which
runs over the Clebsch-multiplicities of symmetric groups. A simple
algorithmic method for determining these multiplicities in terms of
manipulating boxes of Young diagrams, analogous to the one available for
Littlewood-Richardson coefficients, is still an open problem in
symmetric groups. A spacetime interpretation of $\cO^{\L,\mu,R}_{\beta, \tau }$
 might provide stringy insights into this problem.

Many of the techniques in this paper should extend to other gauge
groups. The $SU(N)$ gauge group case could be developed along the
lines of \cite{cr,MKG,brown} and orthogonal and symplectic groups
along the lines of \cite{aabf}. Another generalisation is to consider
operators made from $ X , Y , Z $ along with $ X^{\dagger} ,
Y^{\dagger} , Z^{\dagger} $, and to organise them according to the
degree of singularity in their short distance subtractions. For the
case of $X , X^{\dagger} $ this was done in \cite{kr} using Brauer
algebras. We can also consider the more general protected
operators defined in~\cite{composite}.
  Another interesting line of research is to clarify the
dynamics of strings connected to eighth-BPS giant gravitons in gauge
theory and their connection to space-time geometry.  A lot of progress
in this area has been made for the half-BPS case (see for example for
some of the initial developments and further references
\cite{bbfh,bcv,robi,robii,Bekker:2007ea}).

\vskip.5in 

{ \bf Acknowledgements } We thank Robert de Mello Koch, Francis Dolan,
Yusuke Kimura and Costis Papageorgakis for discussions. SR is
supported by an STFC Advanced Fellowship and in part by the EC Marie
Curie Research Training Network MRTN-CT-2004-512194. PJH is supported
by an EPSRC Standard Research Grant EP/C544250/1.  TWB is on an STFC
studentship.

\begin{appendix}

\section{Formulae}\label{sec:formulae}
The matrices of any representation satisfy the following property,
which follows from Schur's Lemma
\begin{equation}
  \sum_{\s \in S_n} D^R_{ij} (\s )  D^S_{lk} (\s^{-1})  =
  \frac{n!}{d_R}\delta^{RS}  \delta_{ik} \delta_{jl} \label{schurorthogonality}
\end{equation}
We will use orthogonal representation matrices, where orthogonality
guarantees that
\begin{equation}
  D^R_{ij} (\s^{-1}) = D^R_{ji} (\s)
\end{equation}
This means that for orthogonal matrices equation
\eqref{schurorthogonality} becomes
\begin{equation}
  \sum_{\s \in S_n} D^R_{ij} (\s )  D^S_{kl} (\s)  =
  \frac{n!}{d_R}\delta^{RS}  \delta_{ik} \delta_{jl} \label{schurorthogonalityorthogonal}
\end{equation}

\subsection{Clebsch-Gordan coefficients for the tensor product}\label{sec:CGcoef}
The Clebsch-Gordan coefficients allow us to write operators in the
(inner) tensor product space $S \otimes T$, such as $D^S_{ij} (\s)
D^T_{kl} (\s)$, in terms of operators in a single representation $R$.
We obtain them by inserting a complete sets of states
\begin{align}
  D^S_{ij} (\s) D^T_{kl}(\s) & = \bra{S,i;T,k} \s \ket{S,j;T,l} \nn \\ & =
  \sum_{R,R', \tau_R, \tau_{R'}} \braket{S,i;T,k}{R,\tau_R,a}\bra{R,\tau_R,a}
  \s\ket{R',\tau_{R'},b} \braket{R',\tau_{R'},b}{S,j;T,l} \nn \\
& = \sum_{R, \tau_R} \braket{S,i;T,k}{R,\tau_R,a}\; D^R_{ab}(\s)\;
  \braket{R,\tau_R,b}{S,j;T,l} \label{clebschdef}
\end{align}
where the label $\tau_R$ runs over the multiplicity of the appearance of
$R$ in the inner product $ S \otimes T$.  Since the representing
matrices are real, the Clebsch-Gordan coefficients are also real, so
we can write
\begin{equation}\label{clebschoverlap} 
  S^{\tau_R , }{}^{ R}_{ a } \;{}^S_i\;{}^T_k \equiv
  \braket{S,i;T,k}{R,\tau_R,a} =  \braket{R,\tau_R,a}{S,i;T,k}
\end{equation}

The Clebsch-Gordan coefficients satisfy the following orthogonality
relations~\cite{hamermesh}
\begin{align}
 \sum_{j,k}  S^{\tau_R , }{}^{ R}_{ a } \;{}^U_j\;{}^V_k \; 
S^{ \tau_S , }{}^{ S}_{ b} \;{}^U_j
\;{}^V_k & = \delta^{RS} \delta^{ \tau_R \tau_S } 
  \delta_{ab}\label{ortho1} \\
 \sum_{\tau_R }  \sum_R  \sum_a S^{ \tau_R ,}{}^{  R}_{a} \;{}^U_i\;{}^V_j \; 
S^{ \tau_R ,}{}^{ R}_{ a} \;{}^U_k\;{}^V_l & =  \delta_{ik}\delta_{jl} \label{ortho2}
\end{align}
From \eqref{clebschdef} we can then derive
\begin{align}
 \sum_{j,l}  D^S_{ij} (\s) D^T_{kl} (\s)\;S^{ \tau_R,}{}^{ R}_{s}\;{}^S_j\;{}^T_l & =
  \sum_t D^R_{ts} (\s)
  S^{ \tau_R,}{}^{ R}_{ t}\;{}^S_i\;{}^T_k \label{Hamer186} \\
  \sum_\s D^R_{ts} (\s) D^S_{ij} (\s)D^T_{kl} (\s) & = \frac{n!}{d_R} ~ 
 \sum_{\tau_R } 
  S^{ \tau_R ,}{}^{R}_{ t} \;{}^{  S}_i\;{}^T_k \; 
S^{ \tau_R ,}{}^{R}_{s}\;{}^S_j\;{}^T_l 
 \label{sumsigmaid}
\end{align}
Note that, by taking traces in \eqref{sumsigmaid} and using
\eqref{ortho1} we can recover $C(R,S,T)$ \eqref{FHinner} which comes
from the sum over $ \tau_R $.

\section{Some basic linear algebra}\label{sec:some-basic-linear}

We often meet the following simple situation in this paper. We have
a vector space of dimension $d$, $V^d$, sitting inside a vector space of
dimension $D$, $V^D$. We have a projector $P_{ij}$
which projects onto the smaller vector space, ie
\begin{align}
  V^d={v_i:v_i=P_{ij}v_j}\ .
\end{align}
Then if we can write the projector as 
\begin{align}
  P_{ij}=\sum_{\beta=1}^d b_{i\beta} b_{j\beta}\label{decomp}
\end{align}
where 
\begin{align}
\sum_i b_{i\beta} b_{i \beta'}=\delta_{\beta\beta'}\label{bbd}
\end{align}
then the vectors $v_{\beta}:=\sum_i b_{i\beta} v_i$ provide an
orthonormal basis for the subspace $V^d$.

\section{Counting proofs}

\subsection{Explicit gauge-invariant proof}\label{sec:explicit}

The formula for the number of gauge-invariants operators (including
multi-traces) made from fields $\mu_1$ of $X_1$, $\mu_2$ of $X_2$,
\dots $\m_M$ of $X_M$ is given by P\'olya theory to be the coefficient
of $x_1^{\m_1} \cdots x_M^{\m_M}$ in
\begin{align}
&  \prod_{k=1}^\infty \frac{1}{1 - (x_1^k + \cdots +
  x_M^k)}\label{polya} \\
  =& 
  \sum_{i_1, i_2, \dots} (x_1 + \cdots + x_M)^{i_1} \; (x_1^2 + \cdots +
  x_M^2)^{i_2}  \cdots \nn \\
  = & \sum_{i_1, i_2, \dots} \left( \sum_{i_1^1, \dots i_1^M | \sum_\a
  i_1^\a = i_1}\frac{i_1!}{i_1^1 ! \cdots i_1^M!} x_1^{i_1^1} \dots
  x_M^{i_1^M} \right)  \left( \sum_{i_2^1, \dots i_2^M | \sum_\a
  i_2^\a = i_2}\frac{i_2!}{i_2^1 ! \cdots i_2^M!} x_1^{2i_2^1} \dots
  x_M^{2i_2^M} \right)  \cdots \nn \\
  = & \sum_{i_1, i_2, \dots}\;\; \sum_{\{i_1^\a\}, \{i_2^\a\},\dots
  |\sum_\a i_1^\a = i_1, \sum_\a i_2^\a = i_2, \dots}
  \frac{i_1!}{i_1^1 ! \cdots i_1^M!} \frac{i_2!}{i_2^1 ! \cdots
  i_2^M!} \cdots  x_1^{i_1^1 +
  2i_2^1 + \cdots} \cdots x_M^{i_1^M +
  2i_2^M + \cdots} 
\end{align}
In the third line we have used a generalised binomial expansion; in
the fourth line we have collected coefficients of $x_1$, etc.  So we
pick out the coefficient with $\m_1 = i_1^1 + 2i_2^1 + \cdots, \dots
\m_M = i_1^M + 2i_2^M + \cdots$.  This makes the coefficient
\begin{equation}
  N(\m_1, \dots \m_M) = \sum_{\{i_1^\a\}, \{i_2^\a\},\dots
  |\m_k = i_1^k + 2i_2^k + \cdots} \frac{\left(\sum_\a i_1^\a\right)!}{i_1^1 ! \cdots i_1^M!} \frac{\left(\sum_\a i_2^\a\right)!}{i_2^1 ! \cdots \label{yuckyuck}
  i_2^M!} \cdots
\end{equation}

The coefficient $C(R,S,T)$ for $T$ in the inner product for $S_n$
representations $R \otimes S = \sum_T C(R,S,T) T$ is given by
\begin{equation}
  C(R,S,T) = \sum_{{\bf i}} \frac{1}{| \Sym (C_{{\bf i}})| }
  \chi_R(C_{{\bf i}}) \chi_S(C_{{\bf i}}) \chi_T(C_{{\bf i}}) \label{FHinner}
\end{equation}
Notice that it is symmetric in $R,S,T$.  $C_{{\bf i}}$ represents the
conjugacy class of $S_n$ with $i_1$ 1-cycles, $i_2$ 2-cycles, \dots
$i_n$ $n$-cycles (of course we have $\sum_\a \a i_\a = n$, which is
just the condition that $C_{{\bf i}}$ is in $S_n$).  $| \Sym (C_{{\bf
i}})|$ is the size of the symmetry group of the conjugacy class
$C_{{\bf i}}$ and satisfies
\begin{equation}
  | \Sym (C_{{\bf i}})| = i_1 ! 1^{i_1} \cdot  i_2 ! 2^{i_2} \cdots
    i_n ! n^{i_n} = \frac{n!}{|[C_{{\bf i}}]|}
\end{equation}
Here $|[C_{{\bf i}}]|$ is the size of the conjugacy class of $C_{{\bf
i}}$.

We can immediately simplify $\sum_R C(R,R,\L)$ using the orthogonality
relation
\begin{equation}
  \sum_{R} \chi_R (\rho) \chi_R (\tau) = | \Sym(\tau) |
   \;\;\delta([\tau] = [\rho])
\end{equation}
We get
\begin{equation}
  \sum_R C(R,R,\L) = \sum_{C_{\bf i} \in S_n} \chi_{\L}(C_{{\bf i}})
\end{equation}

Now apply these formulae to the expression for  $N(\m_1, \dots \m_M)$
in terms of representation-theoretic data 
\begin{align}
 & N(\m_1, \dots \m_M)\nn\\
  & = \sum_R \sum_{\L} C(R,R,\L)
  g([\m_1], \dots [\m_M]; \L ) \nn \\
  & = \sum_{\L}  \sum_{C_{\bf i}\in S_n} \chi_\L(C_{{\bf i}})
   \frac{1}{\m_1!\cdots \m_M!} \sum_{\r_1\in
    S_{\m_1}}\cdots \sum_{\r_M\in S_{\m_M}} \chi_{[\m_1]}(\r_1) \cdots
  \chi_{[\m_M]}(\rho_M) \chi_\L(\r_1 \circ \cdots \circ \r_M)
\end{align}
Here we have used the formula for Littlewood-Richardson coefficients
\begin{align}
  \label{gRST}
  g(R_1,\dots R_k;T) = 
  \frac{1}{n_1!\cdots n_k!} \sum_{\r_1\in
    S_{n_1}}\cdots \sum_{\r_k\in S_{n_k}} \chi_{R_1}(\r_1) \cdots
  \chi_{R_k}(\rho_k) \chi_T(\r_1 \circ \cdots\circ \r_k)
\end{align}

But of course we know that $\chi_{[\m]}(\r) = 1 \; \forall \r$ and we
can perform the $R$ sum
\begin{align}
  N(\m_1, \dots \m_M) =& \sum_{\L}  \sum_{{\bf i}} \chi_\L(C_{{\bf i}})  \frac{1}{\m_1!\cdots \m_M!} \sum_{\r_1\in
    S_{\m_1}}\cdots \sum_{\r_M\in S_{\m_M}}\chi_\L(\r_1 \circ \cdots \circ
    \r_M) \nn \\
     =  & \frac{1}{\m_1!\cdots \m_M!} \sum_{\r_1\in
    S_{\m_1}}\cdots \sum_{\r_M\in S_{\m_M}}  | \Sym(\r_1 \circ \cdots \circ
    \r_M)|
\end{align}
If we sum over conjugacy classes $C_{{\bf i}^\a}$ in $S_{\m_\a}$
instead we get
\begin{align}
   N(\m_1, \dots \m_M)
     =  & \frac{1}{\m_1!\cdots \m_M!} \sum_{ C_{{\bf i}^1} \in
    S_{\m_1}}\cdots \sum_{ C_{{\bf i}^M}\in S_{\m_M}}  |[C_{{\bf i}^1}] |
     \cdots |[C_{{\bf i}^M}] | | \Sym(C_{{\bf i}^1} \circ \cdots \circ
    C_{{\bf i}^M})| \nn \\
    & = \sum_{ C_{{\bf i}^1} \in
    S_{\m_1}}\cdots \sum_{ C_{{\bf i}^M}\in S_{\m_M}}\frac{| \Sym(C_{{\bf i}^1} \circ \cdots \circ
    C_{{\bf i}^M})|}{| \Sym(C_{{\bf i}^1}) | \cdots  |\Sym(C_{{\bf
     i}^M})|} \nn \\
    & = \sum_{  C_{{\bf i}^1} \in
    S_{\m_1}}\cdots \sum_{ C_{{\bf i}^M}\in S_{\m_M}}\frac{\left(\sum_\a
     i^\a_1 \right)! \cdots \left(\sum_\a
     i^\a_n \right)!  }{  i^1_1 ! \;i^1_2 !\cdots i^1_n !   \cdots
     i^M_1 ! \;i^M_2 !\cdots i^M_n !}  \label{eq:yacountin}
\end{align}
Finally we identify this with the P\'olya theory result
\eqref{yuckyuck}.  Note that the quantity summed here can also be
recognised as the character of the representation of $S_n$ induced
from the trivial representation of the subgroup $S_{\m_1} \times
\cdots S_{\m_M}$
\begin{equation}
  \psi_{\L}(C_{{\bf i}}) =  \sum_{\L} g([\m_1],\dots [\m_M]; \L)
  \chi_\L(C_{{\bf i}})
\end{equation}

\subsection{Example of the counting}

Consider the counting for operators with fields $XXY$.

The relevant $S_3$ outer product is
\begin{equation}
  [2] \circ [1] = [3] \oplus [2,1]
\end{equation}
which gives us
\begin{align}
    N(2,1) & = \sum_R \sum_{\L} C(R,R;\L)
  g([2],[1]; \L ) \nn \\
  & =  \sum_R \left\{ C(R,R;[3])  +  C(R,R;[2,1]) \right\}
\end{align}
The relevant inner products are
\begin{align}
  [3] \otimes [3] & = [3]  \nn \\
  [2,1] \otimes [2,1] & = [3] \oplus [2,1] \oplus [1,1,1]  \nn \\
  [1^3] \otimes [1^3] & = [3]
\end{align}
So we get
\begin{equation}
  N(2,1) = 3 + 1= 4
\end{equation}
This counts correctly for $\tr(XXY)$, $\tr(XX)\tr(Y)$,
$\tr(XY)\tr(X)$ and $\tr(X)\tr(X)\tr(Y)$.

\subsection{Trace counting}\label{sec:tracecounting}

%What is the symmetry group of $\a \in S_n$?
%\begin{equation}
%  \Sym(\a) = \{ \s \in S_n | \a = \s \a \s^{-1} \}
%\end{equation}

The counting formula is give by \eqref{eq:yacountin}
\begin{align}
   N(\m_1, \dots \m_M)
     =  & \frac{1}{\m_1!\cdots \m_M!} \sum_{ C_{{\bf i}^1} \in
    S_{\m_1}}\cdots \sum_{ C_{{\bf i}^M}\in S_{\m_M}}  |[C_{{\bf i}^1}] |
     \cdots |[C_{{\bf i}^M}] | | \Sym(C_{{\bf i}^1} \circ \cdots \circ
    C_{{\bf i}^M})| \nn \\
    & = \frac{1}{|H|} \sum_{h \in H} | \Sym(h)|
\end{align}

We need to count $\tr(\a X_1^{\m_1} \cdots X_M^{\m_M})$ up to the
equivalence relation
\begin{equation}
  \a \sim h^{-1} \a h, \;\; h \in H \label{alphasymm}
\end{equation}

One way to find the number of equivalence classes is to sum over all
$\a$, dividing by the size of the equivalence class of each $\a$
\begin{equation}
  N(\m_1, \dots \m_M) = \sum_{[\a]} 1 = \sum_{\a \in S_n} \frac{1}{|
  [\a] |} = \sum_{\a \in S_n}
  \frac{1}{\textrm{no. of distinct } \b \in S_n | \b = h \a h^{-1}
  \textrm{ for } h \in H }
\end{equation}
Na\"ively we might think that
\begin{equation}
  \left[\textrm{no. of
  distinct } \b \in S_n | \b = h \a
  h^{-1}  \textrm{ for } h \in H \right] = |H|
\end{equation}
We must however be careful: elements in $H$ might also be in the
symmetry group of $\a$ ($\Sym(\a) = \{ \s \in S_n | \a = \s \a \s^{-1}
\}$). To hit unique $\b$'s we must use not $H$ but the coset
$H/[\Sym(\a) \cap H]$
\begin{equation}
   \left[\textrm{no. of
  distinct } \b \in S_n | \b = h \a
  h^{-1}  \textrm{ for } h \in H \right] = |H/[\Sym(\a) \cap H]| =
  \frac{|H|}{|\Sym(\a) \cap H|}
\end{equation}
Thus we get
\begin{align}
  N(\m_1, \dots \m_M) & = \frac{1}{|H|}\sum_{\a \in S_n} |\Sym(\a) \cap
  H| \nn \\
  & = \frac{1}{|H|} \sum_{\a \in S_n} \sum_{h \in H}
  [\textrm{s.t. } \a = h \a h^{-1} ] \nn \\
  & = \frac{1}{|H|}\sum_{h \in H}  \sum_{\a \in S_n} 
  [\textrm{s.t. } h = \a h\a^{-1} ] \nn \\
  & =  \frac{1}{|H|} \sum_{h \in H} | \Sym(h)|
\end{align}

\section{Calculating branching coefficients}\label{sec:branching}

$D_{j_1j_2}^\L(\G)$ projects onto a subspace of the $S_n$
representation $\L$ with dimension $g(\m;\L)$; this subspace is given
by the rows/columns of the matrix $D_{j_1j_2}^\L(\G)$.  We want to
find the branching coefficients $B_{j\b}$ given by
\begin{equation}
  D_{j_1j_2}^\L(\G) =\sum_\b B_{j_1\b}B_{j_2\b}
\end{equation}
We work out some examples below.

\subsection{Highest weight case}

For the highest weight state with $\m = \L$ (for which $g(\m;\L) = 1$)
Hamermesh's basis works such that
\begin{equation}
  D_{j_1j_2}^\L(\G) = \delta_{j_1 1} \delta_{j_2 1}
\end{equation}
Thus the subspace is spanned by a single vector $B_{j} = \delta_{j1}$,
which satisfies all the appropriate properties.

\subsection{All fields different case}

For $\m_1 = 1, \dots \m_M = 1$, i.e. all the fields are different,
then $H = \id$ and $g(\m;\L) = d_\L$
\begin{equation}
  D_{j_1j_2}^\L(\G) = D_{j_1j_2}^\L(\id) = \delta_{j_1 j_2}
\end{equation}
The most obvious basis satisfying the correct properties is $B_{j\b}
= \delta_{j\b}$ (see XYZ example below).

\subsection{$\L = [2,1]$}

\begin{align}
   \frac{1}{|H|}D^{\L = [2,1], \m = XXY} (\G)& =\frac{1}{2}D^{[2,1]} ((1)(2)(3) + (12)(3)) =  \left( \begin{array}{cc} 1 & 0  \\
      0  &  0   \end{array} \right) = \left( \begin{array}{c} 1 \\ 0
      \end{array} \right)\left( \begin{array}{cc} 1 & 0  \end{array}
      \right) \nn \\
   \frac{1}{|H|}D^{\L = [2,1], \m = XYY} (\G)& = \left( \begin{array}{cc} \frac{1}{4} & \frac{\sqrt{3}}{4}  \\
      \frac{\sqrt{3}}{4}  &  \frac{3}{4}   \end{array} \right) = \left( \begin{array}{c} \frac{1}{2} \\ \frac{\sqrt{3}}{2}
      \end{array} \right)\left( \begin{array}{cc} \frac{1}{2} & \frac{\sqrt{3}}{2}  \end{array}
      \right)
\end{align}
Note that for this last one the columns/rows of the matrix aren't
independent (which concurs with the fact that $g=1$), so the subspace
is spanned by the first column say.
\begin{align}
   \frac{1}{|H|}D^{\L = [2,1], \m = XYZ} (\G)& = \left( \begin{array}{cc} 1 & 0  \\
      0  &  1   \end{array} \right) = \left( \begin{array}{c} 1 \\ 0  \end{array} \right)\left( \begin{array}{cc} 1 &0  \end{array}
      \right) +  \left( \begin{array}{c} 0 \\ 1  \end{array} \right)\left( \begin{array}{cc} 0 &1  \end{array}
      \right)
\end{align}

\subsection{$\L = [3,1]$}

\begin{align}
   \frac{1}{|H|}D^{\L = [3,1], \m = XXXY} (\G)&  =  \left(
      \begin{array}{ccc} 1 & 0 & 0 \\
      0  &  0 & 0  \\
      0 & 0 & 0 \end{array} \right) = \left( \begin{array}{c} 1 \\ 0 \\0
      \end{array} \right)\left( \begin{array}{ccc} 1 & 0 & 0 \end{array}
      \right) \nn \\
   \frac{1}{|H|}D^{\L = [3,1], \m = XXYY} (\G)& = \left(
   \begin{array}{ccc} \frac{1}{3} & \frac{\sqrt{2}}{3} & 0 \\
      \frac{\sqrt{2}}{3}  &  \frac{2}{3} & 0 \\
   0 & 0 & 0\end{array} \right) = \left( \begin{array}{c}
   \frac{1}{\sqrt{3}} \\ \frac{\sqrt{2}}{\sqrt{3}} \\ 0 
      \end{array} \right)\left( \begin{array}{ccc} \frac{1}{\sqrt{3}}
   & \frac{\sqrt{2}}{\sqrt{3}} & 0   \end{array}
      \right)
\end{align}
\begin{align}
   \frac{1}{|H|}D^{\L = [3,1], \m = XYYY} (\G)& = \left(
   \begin{array}{ccc} \frac{1}{9} & \frac{\sqrt{2}}{9} & \frac{\sqrt{6}}{9} \\
   \frac{\sqrt{2}}{9}  &  \frac{2}{9} & \frac{2\sqrt{3}}{9} \\
    \frac{\sqrt{6}}{9} & \frac{2\sqrt{3}}{9} & \frac{2}{3}\end{array} \right) = \left( \begin{array}{c}
   \frac{1}{3} \\ \frac{\sqrt{2}}{3} \\ \frac{\sqrt{2}}{\sqrt{3}}
      \end{array} \right)\left( \begin{array}{ccc}
     \frac{1}{3} & \frac{\sqrt{2}}{3} & \frac{\sqrt{2}}{\sqrt{3}}  \end{array}
      \right)
\end{align}
\begin{align}
   \frac{1}{|H|}D^{\L = [3,1], \m = XYYZ} (\G)& = \left(
   \begin{array}{ccc} 1 & 0  & 0 \\
   0  &  \frac{1}{4} & \frac{\sqrt{3}}{4} \\
   0 &  \frac{\sqrt{3}}{4} & \frac{3}{4}\end{array} \right) = \left( \begin{array}{c} 1 \\ 0 \\0
      \end{array} \right)\left( \begin{array}{ccc} 1 & 0 & 0 \end{array}
      \right) +  \left( \begin{array}{c}
   0 \\ \frac{1}{2} \\ \frac{\sqrt{3}}{2}
      \end{array} \right)\left( \begin{array}{ccc}
     0 & \frac{1}{2} & \frac{\sqrt{3}}{2}  \end{array}
      \right) \nn \\
   \frac{1}{|H|}D^{\L = [3,1], \m = XYZZ} (\G)& = \left(
   \begin{array}{ccc} \frac{1}{3} & \frac{\sqrt{2}}{3} & 0 \\
      \frac{\sqrt{2}}{3}  &  \frac{2}{3} & 0 \\
   0 & 0 & 1\end{array} \right) = \left( \begin{array}{c} 0 \\ 0 \\1
      \end{array} \right)\left( \begin{array}{ccc} 0 & 0 & 1 \end{array}
      \right) +\left( \begin{array}{c}
   \frac{1}{\sqrt{3}} \\ \frac{\sqrt{2}}{\sqrt{3}} \\ 0 
      \end{array} \right)\left( \begin{array}{ccc} \frac{1}{\sqrt{3}}
   & \frac{\sqrt{2}}{\sqrt{3}} & 0   \end{array}
      \right) \nn 
\end{align}

\section{Examples of the operators}\label{sec:exampleoperators}

We will work out the operators $\cO^{\Lambda\m , R}_{ \beta, \tau }$
given by \eqref{simplifiedop}

\begin{equation}
    \cO^{ \L\m , R }_{ \b, \tau }
= \frac{1}{n!}\sum_{\a} B_{j \b} \; S^{\tau  ,  \L }_{ \;\;  j
  }\;{}^{R}_{p }\;{}^{R}_{q}\;\; D_{pq}^R(\a) \tr(\a\;
  {\bf X}^\m )
\end{equation}

\subsection{$XY$}

We have
\begin{equation}
  D^{[2]}_{11} (\s) = 1  \quad \quad D^{[1,1]}_{11} (\s) = (-1)^{\s}
  \quad \quad\forall \s
\end{equation}

For $XY$, $\m = [1,1]$, $H = S_1 \times S_1$ and $g([1], [1];[2]) =
g([1],[1];[1,1]) = 1$.  This means there is only one possible value of
$\b$ for each choice of $\L$.  $B_{11} = 1$ for all $\L$.

Since there are no $R$ for which $C(R,R,[1,1])$ is non-zero, we only
have $\L = [2]$.  For each $R$, $C(R,R,[2])=1$, so there is no $\tau$
multiplicity. Thus our two operators are
\begin{align}
  \cO^{\L = [2],R = [2]} &= \frac{1}{2} \sum_{\a} D^{[2]}_{11} (\a)
  \tr(\a\; XY \; ) = \frac{1}{2}[\tr(X)\tr(Y) + \tr(XY)] \nn \\
  \cO^{\L = [2],R = [1,1]} &= \frac{1}{2} \sum_{\a} D^{[1,1]}_{11}
  (\a)  \tr(\a \; XY \;  )  = \frac{1}{2}[ \tr(X)\tr(Y) - \tr(XY)] \nn
\end{align}

\subsection{$XXY$}

We have
\begin{equation}
  D^{[3]}_{11} (\s) = 1  \quad \quad D^{[1,1,1]}_{11} (\s) = (-1)^{\s}
  \quad \quad\forall \s
\end{equation}
and
\begin{align}
&   D^{[2,1]} ((1)(2)(3)) = \left( \begin{array}{cc} 1 & 0 \\ 0 & 1
   \end{array} \right) \quad \quad   
   D^{[2,1]} ((12)) = \left(
   \begin{array}{cc} 1 & 0 \\ 0 & - 1   \end{array} \right)\nn \\
& D^{[2,1]} ((23)) = \left(
   \begin{array}{cc} -\frac{1}{2} & \frac{\sqrt{3}}{2} \\
     \frac{\sqrt{3}}{2} & \frac{1}{2}   \end{array} \right) \quad\quad
 D^{[2,1]} ((13)) = \left(
   \begin{array}{cc} -\frac{1}{2} & -\frac{\sqrt{3}}{2} \\
     -\frac{ \sqrt{3}}{2} & \frac{1}{2}   \end{array} \right)\nn \\
& D^{[2,1]} ((123)) = \left(
   \begin{array}{cc} -\frac{1}{2} & \frac{\sqrt{3}}{2} \\
     -\frac{ \sqrt{3}}{2} & -\frac{1}{2}   \end{array} \right)
     \quad\quad
 D^{[2,1]} ((132)) = \left(
   \begin{array}{cc} -\frac{1}{2} & -\frac{\sqrt{3}}{2} \\
     \frac{ \sqrt{3}}{2} &  - \frac{1}{2}   \end{array} \right)\nn
\end{align}

By the counting there are four possibilities: $\L = [3]$ with
$R=[3],[2,1],[1,1,1]$ and $\L = [2,1]$ with $R = [2,1]$.  For these
possibilities $C(R,R,\L) = 1$ so there is no $\tau$ multiplicity.
$g([2],[1];[3]) = g([2],[1];[2,1]) = 1$ so there is no $\b$ multiplicity.

For $XXY$, $\m = [2,1]$ and $H = S_2 \times S_1$.  For $\L = [3]$
$B_{11} = 1$ and for $\L = [2,1]$ $B_{j1} = \delta_{j1}$, since this
field content is a highest weight state for this $\L$.

We can work out the relevant Clebsch-Gordan coefficients up to a sign
using the identity \eqref{sumsigmaid} when $\tau$ only takes a single
value
\begin{equation}
  \left( S^{\L}_{i}\;{}^{R}_{k}\;{}^{R}_{l} \right)^{2} = \frac{d_\L}{n!} \sum_\s
  D^\L_{ii}(\s)  D^R_{kk}(\s) D^R_{ll}(\s)
\end{equation}
To fix the sign we must work out identities such as \eqref{Hamer186}
explicitly.  We get non-zero values
\begin{equation}
  S^{[3]}_{1}\;{}^{R}_{k}\;{}^{R}_{l}  = \frac{1}{\sqrt{d_R}}
  \delta_{kl} \nn
\end{equation}
\begin{align}
  S^{[2,1]}_{1}\;{}^{[2,1]}_{1}\;{}^{[2,1]}_{1}  & = \frac{1}{\sqrt{2}} \nn\\
  S^{[2,1]}_{1}\;{}^{[2,1]}_{2}\;{}^{[2,1]}_{2}  & = - \frac{1}{\sqrt{2}} \nn
\end{align}

So we have
\begin{align}
  \cO^{\L = [3],R}_{\m = [2,1]} = & \frac{1}{ 3!}  \sum_{\a} \;
  S^{[3]}_{1}\;{}^{R}_{k}\;{}^{R}_{l}  D^{R}_{kl} (\a)
  \tr(\a \; XXY \; )  \nn\\
=&  \frac{1}{3!\sqrt{d_R}}  \sum_{\a} \;
  \chi_{R} (\a)
  \tr(\a \; XXY \; ) \nn
\end{align}
\begin{align}
  \cO^{\L = [2,1],R = [2,1]}_{\m = [2,1]} = & \frac{1}{ 3!}    \sum_{\a} \; 
  S^{[2,1]}_{1}\;{}^{[2,1]}_{k}\;{}^{[2,1]}_{l}  D^{[2,1]}_{kl} (\a)
  \tr(\a \; XXY \;  ) \nn\\
= &  \frac{1}{ 3!}  \sum_{\a} \; \left( \frac{1}{\sqrt{2}} D^{[2,1]}_{11} (\a) -
  \frac{1}{\sqrt{2}} D^{[2,1]}_{22} (\a) \right)
  \tr(\a \; XXY \;  ) \nn\\
 = & \frac{1}{3\sqrt{2}} \left[\tr(Y)\tr(XX)- \tr(X)\tr(XY)   \right] \nn
\end{align}

\subsection{$XYY$ for $\L = [2,1]$}
\begin{align}
    \cO^{\L = [2,1],R = [2,1]}_{\m = [1,2]} & = \sum_j  {\langle} [2,1] ( S_3 ) \rightarrow S_1 \times
S_2 , {\bf 1} | [2,1] , j  {\rangle}  \cO^{\L = [2,1],R = [2,1]}_{j,\m =
  [1,2]} \nn\\
& = \frac{1}{2} \cO^{\L = [2,1],R = [2,1]}_{j=1,\m =
  [1,2]} + \frac{\sqrt{3}}{2} \cO^{\L = [2,1],R = [2,1]}_{j=2,\m =
  [1,2]} \nn \\
& =  \frac{1}{3\sqrt{2}} \left[\tr(Y)\tr(XY) - \tr(X)\tr(YY)  \right] \nn
\end{align}

\subsection{$XXYY$ for $\L = [2,2]$}

We have $C([2,2],[3,1],[3,1]) = C([2,2],[2,2],[2,2]) =
C([2,2],[2,1,1],[2,1,1]) = 1$ and $C([2,2],[5],[5]) =
C([2,2],[1^5],[1^5]) = 0$. This field content gives the highest weight
state for $\L = [2,2]$.  Here $\Phi_r \otimes \Phi^r = X \otimes Y - Y
\otimes X$.
 \begin{equation}
    \cO^{\L = [2,2],R = [2,2]}  = \frac{1}{12\sqrt{2}} \left[\tr(\Phi_r \Phi_s)\tr(\Phi^r)\tr( \Phi^s) +
 \tr(\Phi_r\Phi^r \Phi_s\Phi^s) \right] \nn
\end{equation}
\begin{equation}
    \cO^{\L = [2,2],R = [3,1]}  = \frac{1}{12\sqrt{6}}
    \left[\tr(\Phi_r \Phi_s)\tr(\Phi^r)\tr( \Phi^s) +\tr(\Phi_r
      \Phi_s)\tr(\Phi^r \Phi^s) - 
 \tr(\Phi_r\Phi^r \Phi_s\Phi^s) \right] \nn
\end{equation}
\begin{equation}
    \cO^{\L = [2,2],R = [2,1,1]}  = \frac{1}{12\sqrt{6}}
    \left[\tr(\Phi_r \Phi_s)\tr(\Phi^r)\tr( \Phi^s) - \tr(\Phi_r
      \Phi_s)\tr(\Phi^r \Phi^s) - 
 \tr(\Phi_r\Phi^r \Phi_s\Phi^s) \right] \nn
\end{equation}

\end{appendix}


\begin{thebibliography}{99}

\bibitem{cjr}
  S.~Corley, A.~Jevicki and S.~Ramgoolam,
  ``Exact correlators of giant gravitons from dual N = 4 SYM theory,''
  Adv.\ Theor.\ Math.\ Phys.\  {\bf 5} (2002) 809
  [arXiv:hep-th/0111222].
  %%CITATION = HEP-TH 0111222;%%

\bibitem{malda} 
  J.~M.~Maldacena,
  ``The large N limit of superconformal field theories and supergravity,''
  Adv.\ Theor.\ Math.\ Phys.\  {\bf 2} (1998) 231
  [Int.\ J.\ Theor.\ Phys.\  {\bf 38} (1999) 1113]
  [arXiv:hep-th/9711200];
  %%CITATION = HEP-TH 9711200;%%
  S.~S.~Gubser, I.~R.~Klebanov and A.~M.~Polyakov,
  ``Gauge theory correlators from non-critical string theory,''
  Phys.\ Lett.\ B {\bf 428} (1998) 105
  [arXiv:hep-th/9802109];
  %%CITATION = HEP-TH 9802109;%%
  E.~Witten,
  ``Anti-de Sitter space and holography,''
  Adv.\ Theor.\ Math.\ Phys.\  {\bf 2} (1998) 253
  [arXiv:hep-th/9802150].
  %%CITATION = HEP-TH 9802150;%%

\bibitem{mst}
  J.~McGreevy, L.~Susskind and N.~Toumbas,
  ``Invasion of the giant gravitons from anti-de Sitter space,''
  JHEP {\bf 0006} (2000) 008
  [arXiv:hep-th/0003075].
  %%CITATION = HEP-TH 0003075;%%

\bibitem{giantgravitons}
  M.~T.~Grisaru, R.~C.~Myers and O.~Tafjord,
  ``SUSY and Goliath,''
  JHEP {\bf 0008}, 040 (2000)
  [arXiv:hep-th/0008015];
  %%CITATION = HEP-TH 0008015;%%
  A.~Hashimoto, S.~Hirano and N.~Itzhaki,
  ``Large branes in AdS and their field theory dual,''
  JHEP {\bf 0008}, 051 (2000)
  [arXiv:hep-th/0008016];
  V.~Balasubramanian, M.~Berkooz, A.~Naqvi and M.~J.~Strassler,
  ``Giant gravitons in conformal field theory,''
  JHEP {\bf 0204} (2002) 034
  [arXiv:hep-th/0107119].
  %%CITATION = HEP-TH 0107119;%%


%\cite{Lin:2004nb}
\bibitem{Lin:2004nb}
  H.~Lin, O.~Lunin and J.~M.~Maldacena,
  ``Bubbling AdS space and 1/2 BPS geometries,''
  JHEP {\bf 0410} (2004) 025
  [arXiv:hep-th/0409174].
  %%CITATION = JHEPA,0410,025;%%


\bibitem{zelobenko} 
D.P. Zelobenko, ``Compact Lie Groups and their Representations,'' AMS, 1991.
  

%\cite{fultonharris}
\bibitem{fultonharris}
  W.~Fulton and J.~Harris,
  ``Representation Theory: A First Course,''
  Springer, 1991.

%\cite{hamermesh}
\bibitem{hamermesh}
  M.~ Hamermesh,
  ``Group THeory and its Applications to Physical Problems,''
  Addison-Wesley Publishing Company, 1962

%\cite{somers}
\bibitem{somers} 
  L.J.~Somers,
   ``Analysis of the outer product for the symmetric group,''
  Journal of Mathematical Physics, Volume 24, Issue 4, April 1983, pp.772-778.

%\cite{Dolan:2007rq}
\bibitem{Dolan:2007rq}
  F.~A.~Dolan,
  ``Counting BPS operators in N=4 SYM,''
  arXiv:0704.1038 [hep-th].
  %%CITATION = ARXIV:0704.1038;%%


\bibitem{sundborg}
  B.~Sundborg,
  ``The Hagedorn transition, deconfinement and N = 4 SYM theory,''
  Nucl.\ Phys.\  B {\bf 573} (2000) 349
  [arXiv:hep-th/9908001].
  %%CITATION = NUPHA,B573,349;%%


\bibitem{ammpr}
  O.~Aharony, J.~Marsano, S.~Minwalla, K.~Papadodimas and M.~Van Raamsdonk,
  ``The Hagedorn / deconfinement phase transition in weakly coupled large N
  gauge theories,''
  Adv.\ Theor.\ Math.\ Phys.\  {\bf 8} (2004) 603
  [arXiv:hep-th/0310285].
  %%CITATION = 00203,8,603;%%


\bibitem{cr} 
  S.~Corley and S.~Ramgoolam,
   ``Finite factorization equations and sum rules for BPS correlators in  N = 4
  SYM theory,''
  Nucl.\ Phys.\ B {\bf 641} (2002) 131
  [arXiv:hep-th/0205221].
  %%CITATION = HEP-TH 0205221;%%

%\cite{Bianchi:2006ti}
\bibitem{Bianchi:2006ti}
  M.~Bianchi, F.~A.~Dolan, P.~J.~Heslop and H.~Osborn,
  ``N = 4 superconformal characters and partition functions,''
  Nucl.\ Phys.\  B {\bf 767} (2007) 163
  [arXiv:hep-th/0609179].
  %%CITATION = NUPHA,B767,163;%%

\bibitem{Bars:1983se}
I.~Bars, B.~Morel, and H.~Ruegg, ``Kac-Dynkin diagrams and
supertableaux,''
  {\em J. Math. Phys.} {\bf 24} (1983)
2253.
%%CITATION = JMAPA,24,2253;%%.

\bibitem{turvir}
  V.~G.~Turaev and O.~Y.~Viro,
  ``State sum invariants of 3 manifolds and quantum 6j symbols,''
  Topology {\bf 31} (1992) 865.
  %%CITATION = TPLGA,31,865;%%

\bibitem{cfs}
  S.~w.~Chung, M.~Fukuma and A.~D.~Shapere,
  ``Structure of topological lattice field theories in three-dimensions,''
  Int.\ J.\ Mod.\ Phys.\  A {\bf 9} (1994) 1305
  [arXiv:hep-th/9305080].
  %%CITATION = IMPAE,A9,1305;%%

\bibitem{reshtur}
  N.~Reshetikhin and V.~G.~Turaev,
  ``Invariants of three manifolds via link polynomials and quantum groups,''
  Invent.\ Math.\  {\bf 103} (1991) 547.
  %%CITATION = INVMB,103,547;%%

\bibitem{dijkwit}
  R.~Dijkgraaf and E.~Witten,
  ``Topological Gauge Theories And Group Cohomology,''
  Commun.\ Math.\ Phys.\  {\bf 129} (1990) 393.
  %%CITATION = CMPHA,129,393;%%

\bibitem{baeztaylor}
  J.~Baez and W.~Taylor,
  ``Strings and two-dimensional QCD for finite N,''
  Nucl.\ Phys.\  B {\bf 426} (1994) 53
  [arXiv:hep-th/9401041].
  %%CITATION = NUPHA,B426,53;%%

\bibitem{brown} 
  T.~W.~Brown,
  ``Half-BPS SU(N) correlators in N = 4 SYM,''
  arXiv:hep-th/0703202.
  %%CITATION = HEP-TH/0703202;%%

%\cite{copto}
\bibitem{copto}
  T.~Brown, R.~de Mello Koch, S.~Ramgoolam and N.~Toumbas,
  ``Correlators, probabilities and topologies in N = 4 SYM,''
  JHEP {\bf 0703} (2007) 072
  [arXiv:hep-th/0611290].
  %%CITATION = HEP-TH 0611290;%%


%\cite{D'Hoker:2003vf}
\bibitem{D'Hoker:2003vf}
  E.~D'Hoker, P.~Heslop, P.~Howe and A.~V.~Ryzhov,
  ``Systematics of quarter BPS operators in N = 4 SYM,''
  JHEP {\bf 0304} (2003) 038
  [arXiv:hep-th/0301104].
  %%CITATION = JHEPA,0304,038;%%

%\cite{Ryzhov:2001bp}
\bibitem{Ryzhov:2001bp}
  A.~V.~Ryzhov,
  ``Quarter BPS operators in N = 4 SYM,''
  JHEP {\bf 0111} (2001) 046
  [arXiv:hep-th/0109064].
  %%CITATION = JHEPA,0111,046;%%

\bibitem{Intriligator:1998ig}
  K.~A.~Intriligator,
  ``Bonus symmetries of N = 4 super-Yang-Mills correlation functions via  AdS
  duality,''
  Nucl.\ Phys.\  B {\bf 551} (1999) 575
  [arXiv:hep-th/9811047].
  %%CITATION = NUPHA,B551,575;%%

%\cite{Eden:1999gh}
\bibitem{Eden:1999gh}
  B.~Eden, P.~S.~Howe and P.~C.~West,
  ``Nilpotent invariants in N = 4 SYM,''
  Phys.\ Lett.\  B {\bf 463} (1999) 19
  [arXiv:hep-th/9905085].
  %%CITATION = PHLTA,B463,19;%%

%\cite{Heslop:2001gp}
\bibitem{3pnt}
  P.~J.~Heslop and P.~S.~Howe,
  ``OPEs and 3-point correlators of protected operators in N = 4 SYM,''
  Nucl.\ Phys.\  B {\bf 626} (2002) 265
  [arXiv:hep-th/0107212].
  %%CITATION = NUPHA,B626,265;%%

\bibitem{sym}
  P.~J.~Heslop and P.~S.~Howe,
  ``Aspects of N = 4 SYM,''
  JHEP {\bf 0401} (2004) 058
  [arXiv:hep-th/0307210].
  %%CITATION = JHEPA,0401,058;%%

\bibitem{mik} 
  A.~Mikhailov,
  ``Giant gravitons from holomorphic surfaces,''
  JHEP {\bf 0011} (2000) 027
  [arXiv:hep-th/0010206].
  %%CITATION = JHEPA,0011,027;%%

\bibitem{beasley}
  C.~E.~Beasley,
  ``BPS branes from baryons,''
  JHEP {\bf 0211} (2002) 015
  [arXiv:hep-th/0207125].
  %%CITATION = JHEPA,0211,015;%%


\bibitem{bglm}
  I.~Biswas, D.~Gaiotto, S.~Lahiri and S.~Minwalla,
  ``Supersymmetric states of N = 4 Yang-Mills from giant gravitons,''
  [arXiv:hep-th/0606087].
  %%CITATION = HEP-TH/0606087;%%

\bibitem{mansur}
  G.~Mandal and N.~V.~Suryanarayana,
  ``Counting 1/8-BPS dual-giants,''
  JHEP {\bf 0703} (2007) 031
  [arXiv:hep-th/0606088].
  %%CITATION = JHEPA,0703,031;%%

\bibitem{chenetal}
  B.~Chen {\it et al.},
  ``Bubbling AdS and droplet descriptions of BPS geometries in IIB
  supergravity,''
  arXiv:0704.2233 [hep-th].
  %%CITATION = ARXIV:0704.2233;%%

\bibitem{gmno}
  E.~Gava, G.~Milanesi, K.~S.~Narain and M.~O'Loughlin,
  ``1/8 BPS states in AdS/CFT,''
  JHEP {\bf 0705} (2007) 030
  [arXiv:hep-th/0611065].
  %%CITATION = JHEPA,0705,030;%%

\bibitem{skentay} 
  K.~Skenderis and M.~Taylor,
  ``Anatomy of bubbling solutions,''
  JHEP {\bf 0709} (2007) 019
  [arXiv:0706.0216 [hep-th]].
  %%CITATION = JHEPA,0709,019;%%

\bibitem{sunborg}
B.~Sundborg,
  ``Stringy gravity, interacting tensionless strings and massless higher
  spins,''
  Nucl.\ Phys.\ Proc.\ Suppl.\  {\bf 102} (2001) 113
  [arXiv:hep-th/0103247].
  %%CITATION = HEP-TH 0103247;%%

\bibitem{sezgin}
  E.~Sezgin and P.~Sundell,
  ``Massless higher spins and holography,''
  Nucl.\ Phys.\ B {\bf 644} (2002) 303
  [Erratum-ibid.\ B {\bf 660} (2003) 403]
  [arXiv:hep-th/0205131].
  %%CITATION = HEP-TH 0205131;%%

\bibitem{bianchi}
  M.~Bianchi, J.~F.~Morales and H.~Samtleben,
  ``On stringy AdS5 x $S^5$ and higher spin holography,''
  JHEP {\bf 0307} (2003) 062
  [arXiv:hep-th/0305052].
  %%CITATION = HEP-TH 0305052;%%

%\cite{bouffe}
\bibitem{bouffe}
  M.~Bianchi, P.~J.~Heslop and F.~Riccioni,
  ``More on la grande bouffe,''
  JHEP {\bf 0508} (2005) 088
  [arXiv:hep-th/0504156].
  %%CITATION = JHEPA,0508,088;%%

\bibitem{taktsu}
  Y.~Takayama and A.~Tsuchiya,
  ``Complex matrix model and fermion phase space for bubbling AdS
  geometries,''
  JHEP {\bf 0510} (2005) 004
  [arXiv:hep-th/0507070].
  %%CITATION = JHEPA,0510,004;%%

\bibitem{djr}
  A.~Donos, A.~Jevicki and J.~P.~Rodrigues,
  ``Matrix model maps in AdS/CFT,''
  Phys.\ Rev.\  D {\bf 72} (2005) 125009
  [arXiv:hep-th/0507124].
  %%CITATION = PHRVA,D72,125009;%%

\bibitem{emergent} 
  D.~Berenstein,
  ``Large N BPS states and emergent quantum gravity,''
  JHEP {\bf 0601} (2006) 125
  [arXiv:hep-th/0507203].
  %%CITATION = JHEPA,0601,125;%%

\bibitem{bcls}
  V.~Balasubramanian, B.~Czech, K.~Larjo and J.~Simon,
  ``Integrability vs. information loss: A simple example,''
  JHEP {\bf 0611} (2006) 001
  [arXiv:hep-th/0602263].
  %%CITATION = JHEPA,0611,001;%%  

\bibitem{agpol} 
  A.~Agarwal and A.~P.~Polychronakos,
  ``BPS operators in N = 4 SYM: Calogero models and 2D fermions,''
  JHEP {\bf 0608} (2006) 034
  [arXiv:hep-th/0602049].
  %%CITATION = JHEPA,0608,034;%%

\bibitem{malstrom} 
  J.~M.~Maldacena and A.~Strominger,
  ``AdS(3) black holes and a stringy exclusion principle,''
  JHEP {\bf 9812} (1998) 005
  [arXiv:hep-th/9804085].
  %%CITATION = JHEPA,9812,005;%%

\bibitem{MKG}
  R.~de Mello Koch and R.~Gwyn,
  ``Giant graviton correlators from dual SU(N) super Yang-Mills theory,''
  JHEP {\bf 0411} (2004) 081
  [arXiv:hep-th/0410236].
  %%CITATION = JHEPA,0411,081;%%

\bibitem{aabf}
  O.~Aharony, Y.~E.~Antebi, M.~Berkooz and R.~Fishman,
  ``'Holey sheets': Pfaffians and subdeterminants as D-brane
 operators in large
  N gauge theories,''
  JHEP {\bf 0212} (2002) 069
  [arXiv:hep-th/0211152].
  %%CITATION = JHEPA,0212,069;%%

\bibitem{kr} 
  Y.~Kimura and S.~Ramgoolam,
  ``Branes, Anti-Branes and Brauer Algebras in Gauge-Gravity duality,''
  arXiv:0709.2158 [hep-th].
  %%CITATION = ARXIV:0709.2158;%%

%\cite{Heslop:2001dr}
\bibitem{composite}
  P.~J.~Heslop and P.~S.~Howe,
  ``A note on composite operators in N = 4 SYM,''
  Phys.\ Lett.\  B {\bf 516} (2001) 367
  [arXiv:hep-th/0106238].
  %%CITATION = PHLTA,B516,367;%%

\bibitem{bbfh}
  V.~Balasubramanian, D.~Berenstein, B.~Feng and M.~x.~Huang,
  ``D-branes in Yang-Mills theory and emergent gauge symmetry,''
  JHEP {\bf 0503} (2005) 006
  [arXiv:hep-th/0411205].
  %%CITATION = JHEPA,0503,006;%%

\bibitem{bcv}
  D.~Berenstein, D.~H.~Correa and S.~E.~Vazquez,
  ``A study of open strings ending on giant gravitons, spin chains and
  integrability,''
  JHEP {\bf 0609} (2006) 065
  [arXiv:hep-th/0604123].
  %%CITATION = JHEPA,0609,065;%%

\bibitem{robi}
  R.~de Mello Koch, J.~Smolic and M.~Smolic,
  ``Giant Gravitons - with Strings Attached (I),''
  JHEP {\bf 0706} (2007) 074
  [arXiv:hep-th/0701066].
  %%CITATION = JHEPA,0706,074;%%

\bibitem{robii}
  R.~de Mello Koch, J.~Smolic and M.~Smolic,
  ``Giant Gravitons - with Strings Attached (II),''
  JHEP {\bf 0709} (2007) 049
  [arXiv:hep-th/0701067].
  %%CITATION = JHEPA,0709,049;%%

%\cite{Bekker:2007ea}
\bibitem{Bekker:2007ea}
  D.~Bekker, R.~de Mello Koch and M.~Stephanou,
  ``Giant Gravitons - with Strings Attached (III),''
  arXiv:0710.5372 [hep-th].
  %%CITATION = ARXIV:0710.5372;%%


\end{thebibliography}
\end{document}